\documentclass[pra,aps,twocolumn,nopacs,superscriptaddress,nofootinbib]{revtex4}

\usepackage{graphicx}  
\usepackage{dcolumn}  
\usepackage{bm}           
\usepackage{bbm}
\usepackage{amsmath}
\usepackage{epsfig}
\usepackage{indentfirst}
\usepackage{psfrag}
\usepackage{subfigure}
\usepackage{amssymb}
\usepackage{color}
\usepackage[colorlinks,linkcolor=blue,citecolor=blue,urlcolor=blue,hyperindex,driverfallback=dvipdfm]{hyperref}
\usepackage[T1]{fontenc}

\def\ii{{\rm i}}  \def\ee{{\rm e}}
\def\rb{{\bf r}}      \def\vb{{\bf v}}
      \def\nn{\hat{\bf n}}
\def\kb{{\bf k}}    
    
\def\me{m_{\rm e}}  \def\kB{{k_{\rm B}}}
\def\Eb{{\bf E}}    \def\Ab{{\bf A}}
\def\pb{{\bf p}}  
  
        \def\wp{\omega_{\rm p}}

\def\epsilonb{\epsilon_{\rm b}}  \def\fb{{\bf f}}

\begin{document}
\title{Probing Quantum Optical Excitations with Fast Electrons}
\author{Valerio~Di~Giulio}
\affiliation{ICFO-Institut de Ciencies Fotoniques, The Barcelona Institute of Science and Technology, 08860 Castelldefels (Barcelona), Spain}
\author{Mathieu~Kociak}
\affiliation{Laboratoire de Physique des Solides, CNRS, Universit\'{e} Paris Sud XI, F 91405 Orsay, France}
\author{F.~Javier~Garc\'{\i}a~de~Abajo}
\email[Corresponding author: ]{javier.garciadeabajo@nanophotonics.es}
\affiliation{ICFO-Institut de Ciencies Fotoniques, The Barcelona Institute of Science and Technology, 08860 Castelldefels (Barcelona), Spain}
\affiliation{ICREA-Instituci\'o Catalana de Recerca i Estudis Avan\c{c}ats, Passeig Llu\'{\i}s Companys 23, 08010 Barcelona, Spain}

\begin{abstract}
Probing optical excitations with nanometer resolution is important for understanding their dynamics and interactions down to the atomic scale. Electron microscopes currently offer the unparalleled ability of rendering spatially-resolved electron spectra with combined meV and sub-nm resolution, while the use of ultrafast optical pulses enables fs temporal resolution and exposure of the electrons to ultraintense confined optical fields. Here, we theoretically investigate fundamental aspects of the interaction of fast electrons with localized optical modes that are made possible by these advances. We use a quantum-optics description of the optical field to predict that the resulting electron spectra strongly depend on the statistics of the sample excitations (bosonic or fermionic) and their population (Fock, coherent, or thermal), whose autocorrelation functions are directly retrieved from the ratios of electron gain intensities. We further explore feasible experimental scenarios to probe the quantum characteristics of the sampled excitations and their populations.
\end{abstract}
\date{\today}
\maketitle

\section{Introduction} 

Electron energy-loss spectroscopy (EELS) performed in electron microscopes is a fertile source of information on the dielectric properties of materials down to the nanometer scale \cite{H03,MBM07,KLD14}. This technique is widely used to identify chemical species with atomic resolution through their characteristic high-energy core losses \cite{MKM08,ZRB12}. Additionally, low-loss EELS provides insight into the spatial and spectral distributions of plasmons in metallic nanostructures \cite{paper149,RB13,KS14,HTH15,GBL17}, and more recently also of phonons in polaritonic materials \cite{KLD14,LTHB17} thanks to remarkable advances in instrument resolution. In a parallel effort, the ultrafast dynamics of nanostructured materials and their influence on optical near-fields can be studied by synchronizing the arrivals of fs light and electron pulses at the sample \cite{BFZ09,RB16,KML17}. Indeed, although photons and electrons interact extremely weakly in free space due to the lack of energy-momentum matching, the evanescent field components produced upon light scattering by material structures breaks the mismatch, giving rise to efficient light-electron interaction, and effectively producing exchanges of multiple quanta between the electron and the optical field, accompanied by a complex sub-fs dynamics \cite{BFZ09,paper151,PLZ10,KGK14,PLQ15,FES15,paper282,EFS16,FBR17,PRY17,paper306,paper311,paper312,MB18}. Based on this principle, photon-induced near-field electron microscopy (PINEM) is performed by analyzing the resulting multiple gain and loss features in the electron spectra.

PINEM experiments have so far relied on coherent light sources (i.e., lasers), for which the measured spectra are well reproduced by assuming sample bosonic excitations that are coherently populated with a large number of quanta. The probability of each electron spectral peak associated with a net exchange of $\ell$ quanta is then simply given by the squared Bessel function $J_\ell^2(2|\beta|)$, where a single parameter $\beta=(e/\hbar\omega)\int dz\,\mathcal{E}_z(z)\ee^{-\ii\omega z/v}$ captures the strength of the electron-light interaction, mediated by the optical electric field component $\mathcal{E}_z(z)$ along the direction of electron propagation $z$ for an electron velocity $v$ and photon frequency $\omega$ \cite{paper311}. For nanometer-sized samples (e.g., $\Delta z\sim100\,$nm) illuminated at optical frequencies ($\hbar\omega\sim1\,$eV), a field amplitude $\mathcal{E}\sim10^7\,$V/m renders $|\beta|\sim e\Delta z\mathcal{E}/\hbar\omega\sim1$. Eventually, even the zero loss peak (ZLP, corresponding to $\ell=0$) is fully depleted for $|\beta|\approx1.2$. The underlying physics is thus described in terms of a classical optical field interacting with the electron through sample-mediated harmonic evanescent fields. However, we expect new physics to arise when departing from this regime by considering anharmonic states of the illuminated sample, such as those associated with fermionic excitations \cite{TK13,MTC15,BMT16} or when the external light source is not in a coherent state such as that furnished by a laser. As an interesting avenue in this direction, electron-photon entanglement has been recently predicted to influence the interaction with an electron when the sample is previously excited by a trailing electron \cite{K19}.

Here, we discuss the interaction of electron beams with individual optical modes and predict nontrivial characteristics of this interaction when the modes are excited through external illumination depending on the mode nature and population statistics. Specifically, the electron spectra resulting from the interaction with bosonic and fermionic excitations exhibit a radically different dependence on the external light intensity. Additionally, the electron spectra for bosonic modes depend dramatically on the photon statistics, giving rise to a varied phenomenology of asymmetric gain and loss peaks at low fluences and distinct intensity dependences under strong pumping. Interestingly, the autocorrelation functions can be directly retrieved from ratios of measured electron gain intensities. We further propose a feasible experimental realization of these ideas based on a sample consisting of an optical cavity that is fed by optically pumped three-level quantum emitters (QEs).

\section{Interaction between a beam electron and a bosonic excitation}

We consider a sample characterized by a single boson mode of frequency $\omega_0$ interacting with a focused beam electron of momentum and kinetic energy tightly peaked around $\hbar\kb_0$ and $E_0$, respectively. Assuming the sample to have an extension along the beam direction sufficiently small as to preserve the transversal beam profile in the sample region, we can write the incident electron wave function as $\psi_0(\rb,t)=\ee^{\ii(\kb_0\cdot\rb-E_0t/\hbar)}\phi_0(\rb-\vb t)$, where $\phi_0$ is a slowly varying function of the moving-frame position $\rb-\vb t$. Further adopting the nonrecoil approximation ($\hbar\omega_0\ll E_0$) and neglecting inelastic boson losses ($>$ps lifetimes) during the interaction time (in the fs range), we can write the system Hamiltonian $\hat{\mathcal{H}}=\hat{\mathcal{H}}_0+\hat{\mathcal{H}}_1$ as the sum of unperturbed and interaction parts (see Appendix\ \ref{interaction} for a self-contained derivation from the Dirac equation)
\begin{align}
\hat{\mathcal{H}}_0&=\hbar\omega_0 \hat{a}^\dagger \hat{a} + E_0-\hbar \vb\cdot(\ii\nabla+\kb_0), \nonumber\\
\hat{\mathcal{H}}_1&=(e\vb/c)\cdot\hat\Ab,
\nonumber
\end{align}
where $\hat{a}^\dagger$ and $\hat{a}$ are creation and annihilation operators of the boson mode satisfying the commutation relations of a quantum harmonic oscillator, $\vb=(\hbar\kb_0/\me)/(1+E_0/\me c^2)$ is the electron velocity, and $\hat{\mathcal{H}}_1$ describes the minimal coupling between the electron and the electromagnetic vector potential $\hat\Ab$ associated with the sample excitations. This formalism is similar to previous works \cite{paper151,paper272}, with the important difference that we incorporate a free-electron boson term and the vector potential now becomes an operator,
\begin{align}
\hat\Ab=(-\ii c/\omega_0)\left[\vec{\mathcal{E}}_0(\rb)\hat{a}-\vec{\mathcal{E}}^*_0(\rb)\hat{a}^\dagger\right],
\nonumber
\end{align}
where $\vec{\mathcal{E}}_0(\rb)$ is the single-mode electric field amplitude. Upon inspection, taking $\vb$ along $z$, we find the wave function of the sample-electron system to admit the solution
\begin{align}
|\psi(\rb,t)\rangle\!=\!\psi_0(\rb,t)\!\!\!\sum_{\ell=-\infty}^\infty \sum_{n=0}^\infty\ee^{\ii\omega_0[\ell(z/v-t)-nt]}f_\ell^n(z)|n\rangle,
\label{psi}
\end{align}
where $f_\ell^n$ represents the amplitude of the boson Fock state $|n\rangle$ combined with a change $\ell\hbar\omega_0$ in electron energy. Inserting  Eq.\ (\ref{psi}) into the Schr\"odinger equation $\hat{\mathcal{H}}|\psi\rangle=\ii\hbar\partial|\psi\rangle/\partial t$, we find that it is indeed a solution, provided the amplitudes satisfy the equation
\begin{align}
\frac{df_\ell^n}{dz}=\sqrt{n}\,u^*\,f_{\ell+1}^{n-1}-\sqrt{n+1}\,u\,f_{\ell-1}^{n+1},
\label{fln}
\end{align}
where $u(z)=(e/\hbar\omega_0)\mathcal{E}_{0z}(z)\ee^{-\ii\omega_0z/v}$. Interestingly, this expression guarantees that $n+\ell$ is conserved along the interaction (i.e., the number of excitations in the electron-boson system is preserved), which in turn allows us to treat each initial population $p_n$ of the boson state $|n\rangle$ as an independent channel. However, if we are just interested in the transmitted electron spectrum, we can dismiss the relative phases of these channels and initialize the amplitudes as $f_\ell^n(-\infty)=\delta_{\ell0}\sqrt{p_n}$, with the electron prepared in the incident state $\ell=0$. After propagation according to Eq.\ (\ref{fln}), the transmitted EELS probability reduces to $\Gamma(\omega)=\sum_{\ell=-\infty}^\infty P_\ell\,\delta(\omega+\ell\omega_0)$, where
\begin{align}
P_\ell=\sum_{n={\rm max}\{0,-\ell\}}^\infty \left|f_\ell^n(\infty)\right|^2
\label{Pl}
\end{align}
is the probability for the electron to change its energy by $\ell\hbar\omega_0$. Noticing that Eq.\ (\ref{fln}) is formally equivalent to the Schr\"odinger equation for a classically driven quantum oscillator (see Appendix\ \ref{fanalytical}), we find the analytical solution \cite{G1963,LKB1970}
\begin{align}
f_\ell^n(\infty)=&\sqrt{p_{n+\ell}}\,\ee^{\ii\chi}\sqrt{(n+\ell)!n!}\;\ee^{-|\beta_0|^2/2}(-\beta_0)^\ell \nonumber\\
&\times\sum_{n'} \frac{\left(-|\beta_0|^2\right)^{n'}}{n'!(\ell+n')!(n-n')!},
\nonumber
\end{align}
where the sum is limited to the range ${\rm max}\{0,-\ell\}\le n'\le n$, $\chi$ is a global phase that is irrelevant in this study, and
\begin{align}
\beta_0=\frac{e}{\hbar\omega_0}\int dz\,\mathcal{E}_{0z}\ee^{-\ii\omega_0 z/v}
\label{beta1}
\end{align}
is an electron-boson coupling coefficient (like the PINEM $\beta$ coefficient, but with the electric field normalized to one quantum); this result, which is in excellent agreement with direct numerical integration of Eq.\ (\ref{fln}), shows that the interaction depends exclusively on the initial mode population $p_n$ and the parameter $\beta_0$ defined by Eq.\ (\ref{beta1}).

\begin{figure*}
\begin{centering}
\includegraphics[width=0.8\textwidth]{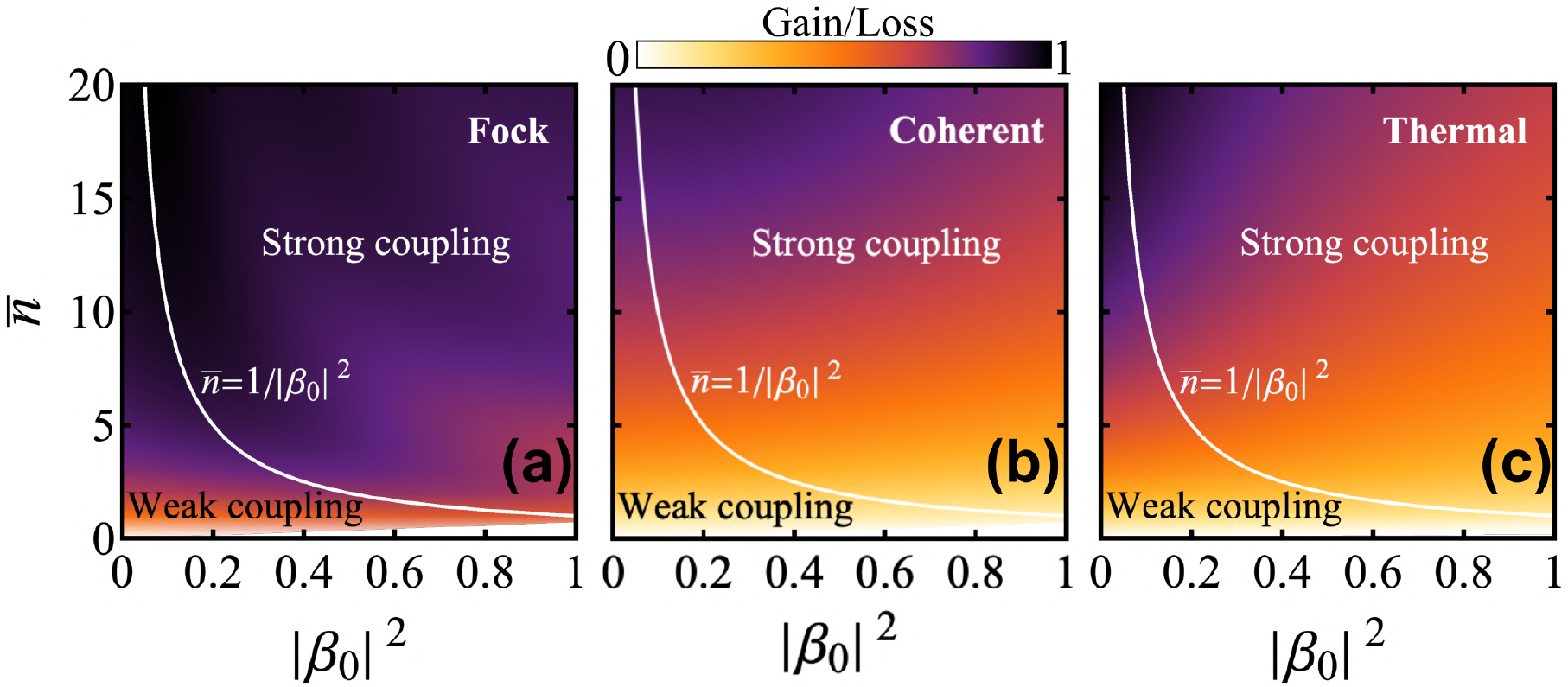}
\par\end{centering}
\caption{Coupling regimes in the interaction of a beam electron with an optical mode. Weak and strong coupling corresponds to the regions roughly separated by the contour $\bar{n}|\beta_0|^2\sim1$ (white line), where $\bar{n}$ is the average mode population and $\beta_0$ is the single-mode interaction coefficient [Eq.\ (\ref{beta1})]. The density plot shows the ratio of integrated gains and losses in the electron spectra for Fock (a), coherent (b), and thermal (c) populations.}
\label{Fig1}
\end{figure*}

\section{Weak coupling limit}

The first-order perturbation solution of Eq.\ (\ref{fln}) readily leads to loss and gain probabilities (see Appendix\ \ref{smallbeta}) $P_{-1}=(\bar{n}+1)|\beta_0|^2$ and $P_1=\bar{n}|\beta_0|^2$, respectively, where $\bar{n}=\sum_{n=1}^\infty n p_n$ is the average mode population and $\beta_0$ is given by Eq.\ (\ref{beta1}). We then find that both loss and gain peaks increase in strength with $|\beta_0|^2$, but their difference is independent of $\bar{n}$. In fact the electron-boson interaction strength is determined by $\bar{n}|\beta_0|^2$, which allows us to separate the regimes of weak and strong coupling depending on $|\beta_0|^2$ and $\bar{n}$, as shown in Fig.\ \ref{Fig1}. Additionally, we observe that the ratio of gains to losses approaches 1 in the $\bar{n}\gg1$ limit, which is consistent with the behavior of the weak-coupling ratio $\bar{n}/(\bar{n}+1)$.

Continuing the  perturbation series, we find that the leading order contribution to the $\ell>0$ gain peak is directly proportional to the $\ell^{\rm th}$-order autocorrelation function at zero delay ${\rm g}^{(\ell)}=\sum_{n=\ell}^\infty n(n-1)\dots(n-\ell+1)p_n/\bar{n}^2$ \cite{L1983}; we find the powerful result $P_\ell/P_1^\ell={\rm g}^{(\ell)}/(\ell!)^2$ (see Appendix\ \ref{smallbeta}), which allows us to extract the autocorrelation functions from ratios of measured gain intensities.

Incidentally, for a fermionic sample mode (e.g., a two-level system) under external cw illumination, we have $f_\ell^{n>1}=0$, $p_1=\bar n$, and $p_0=1-\bar n$, which, to first-order perturbation in the electron-fermion interaction, readily lead to $P_{-1}=(1-\bar{n})|\beta_0|^2$and $P_1=\bar{n}|\beta_0|^2$. As the light intensity increases, the steady-state populations of a lossy fermion approach $p_0=p_1=1/2$ (see Appendix\ \ref{weakcoupling}), leading to loss and gain peaks of equal strength.

\section{Interaction with a dipolar excitation}

For completeness, we discuss a mode described by a transition dipole $\pb$ at the origin, so that the single-mode electric field reduces to $\vec{\mathcal{E}}_0=\left(\pb k_0^2+\pb\cdot\nabla\nabla\right)\ee^{\ii k_0r}/r$, where $k_0=\omega_0/c$, $r=\sqrt{b^2+z^2}$, and $b$ is the electron impact-parameter. This is an excellent approximation for plasmonic particles and Mie resonators in a spectral region dominated by a dipolar excitation \cite{paper149}. Inserting this field into Eq.\ (\ref{beta1}), we find the coupling parameter $\beta_0=(-2e\omega_0/\hbar v^2\gamma)\left[\ii p_x K_1(\zeta)+(p_z/\gamma)K_0(\zeta)\right]$, where the modified Bessel functions $K_m$ are evaluated at $\zeta=\omega_0b/v\gamma$, we introduce the relativistic Lorentz factor $\gamma=1/\sqrt{1-v^2/c^2}$, and the electron beam is taken to move along $z$ and cross the $x$ axis. Reassuringly, the resulting lowest-order loss probability for the ground state sample $|\beta_0|^2$ coincides with the integrated EELS probability for a dipolar scatterer (see  Appendix\ \ref{dipolarEELS}), thus corroborating that the expression assumed for $\vec{\mathcal{E}}_0$ has the correct single-dipole-mode normalization. The values of $\beta_0$ considered in this study are in the range estimated via this model for the dipolar response of metallic (plasmonic) and dielectric (Mie resonances) optical cavities interacting with few to 100s keV electrons (see Appendices\ \ref{plasmoncavity} and \ref{Miecavity}).

\begin{figure*}[t]
\begin{centering}
\includegraphics[width=1.0\textwidth]{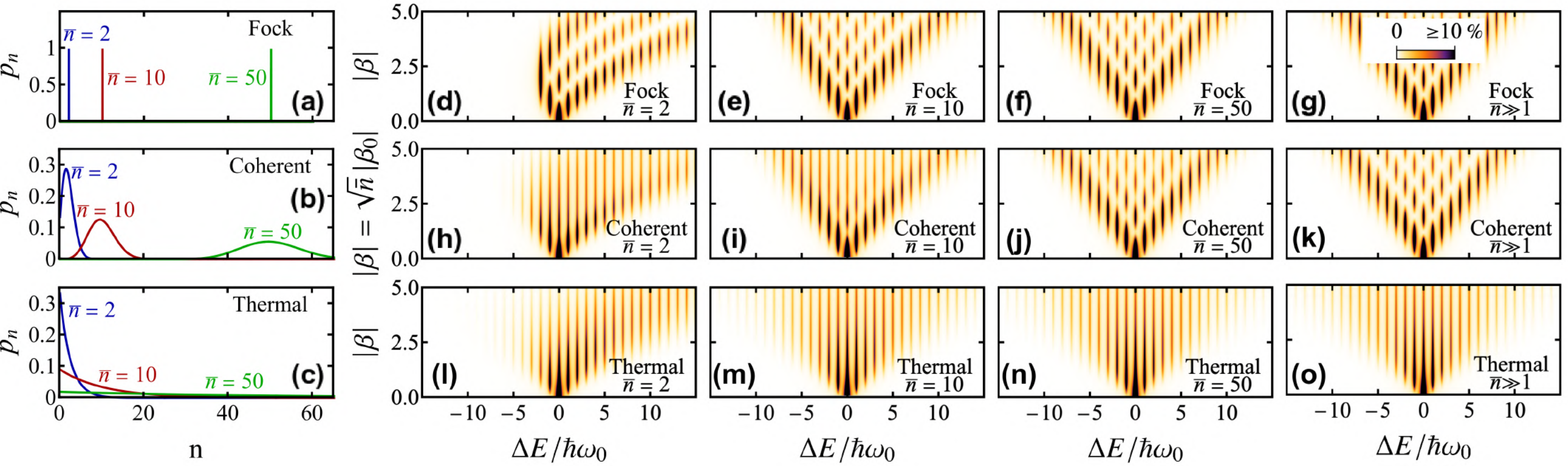}
\par\end{centering}
\caption{Dependence on boson population distribution in the interaction with an electron beam. (a-c) Distribution of the probability $p_n$ for occupation of each state $|n\rangle$ in the three types of population statistics considered at the moment of electron interaction: Fock (a), coherent (b), and thermal (c), with average values $\bar{n}=$2, 10, and 50. (d-o) Electron spectra after interaction with a dipolar mode with the initial populations of (a-c) as a function of the electron-mode coupling parameter $\sqrt{\bar{n}}|\beta_0|$. The energy loss $\Delta E$ is normalized to the boson energy $\hbar\omega_0$ and a peak Lorentzian broadening of $0.1\,\omega_0$ is introduced for clarity.}
\label{Fig2}
\end{figure*}

\section{Dependence on boson population statistics}

The boson mode can be populated in different ways, and in particular when pumped by external illumination, the photon statistics is directly imprinted on it. Here, we study three representative distributions with average occupation $\bar{n}$: (1) a Fock state described by $p_n=\delta_{n,\bar{n}}$, as one would obtain by coupling to quantum emitters (see below); (2) a coherent state characterized by a Poissonian distribution $p_n=\ee^{-\bar{n}}\bar{n}^n/n!$ \cite{G1963}, as provided by external laser illumination; and (3) a thermal distribution, as produced by illuminating with chaotic light or by heating the sample mode at a temperature $T$ commensurate with $\hbar\omega_0/\kB$. The latter is characterized by $p_n=\left(1-\ee^{-\theta}\right)\,\ee^{-n\theta}$, corresponding to a Bose-Einstein average occupation $\bar{n}=1/\left(\ee^\theta-1\right)$, where $\theta=\hbar\omega_0/\kB T$. Examples of these distributions are plotted in Fig.\ \ref{Fig2}(a-c) for $\bar{n}=$2, 10, and 50. We further present in Fig.\ \ref{Fig2}(d-o) the evolution of the transmitted electron spectra for each of the distributions as a function of the coupling parameter $\sqrt{\bar{n}}|\beta_0|$ (vertical scales). The spectra become more asymmetric for smaller $\bar{n}$ because the number of gains cannot substantially exceed $\bar{n}$ (see also Fig.\ \ref{Fig1}), as observed in recent experiments \cite{paper325}, while the number of losses increases indefinitely with $|\beta_0|$.

In contrast, in the $\bar{n}\gg1$ limit, the electron spectra become symmetric with respect to $\ell=0$ [Fig.\ \ref{Fig2}(g,k,o)] when $|\ell|\ll\bar{n}$. We can then approximate the square roots of Eq.\ (\ref{fln}) by $\sqrt{n+\ell}$, leading to the same equation as for PINEM \cite{PLZ10,PZ12,paper272,paper311}, whose solution reads (see Appendix\ \ref{largen})
\begin{align}
f_\ell^n(\infty)=\sqrt{p_{n+\ell}}\,\ee^{\ii\ell\arg\{-\beta_0\}}J_\ell\left(2\sqrt{n+\ell}|\beta_0|\right),
\nonumber
\end{align}
where the prefactor $\sqrt{p_{n+\ell}}$ denotes the initial population of the cavity mode before interaction. Inserting this into Eq.\ (\ref{Pl}) and taking the $\bar{n}\gg1$ limit, we find the analytical expressions (see Appendix\ \ref{largen})
\begin{align}
P_\ell=\left\{\begin{array}{ll}
J_\ell^2(2|\beta|), & \quad\quad \text{Fock, coherent} \\
\ee^{-2|\beta|^2}I_\ell(2|\beta|^2), & \quad\quad \text{thermal}
\end{array} \right.
\label{largenP}
\end{align}
depending on the statistics of the mode population, where $\beta=\sqrt{\bar{n}}\beta_0$. For Fock and coherent distributions, this expression coincides with the well-known PINEM probability \cite{PLZ10,paper272,paper311} for an optical field amplitude $\mathcal{E}_{z}=\sqrt{\bar{n}}\mathcal{E}_{0z}$; the resulting spectra [Fig.\ \ref{Fig2}(g,k)] present the predicted \cite{paper151} and subsequently measured \cite{FES15} quantum-billiard oscillatory structure as a function of both $\ell$ and field strength, as previously studied for model multilevel atoms \cite{ESB1977}. In contrast, a thermal boson distribution leads to a monotonic decrease with increasing $\ell$ and $|\mathcal{E}_{z}|$ [Fig.\ \ref{Fig2}(n,o)]. In all cases, we find an average $\langle|\ell|\rangle\sim|\beta|$. The $\bar{n}\gg1$ limit [Eq.\ (\ref{largenP})] is nearly reached under the conditions of Fig.\ \ref{Fig2} for $\bar{n}=50$ (cf. two rightmost panel columns).

\begin{figure*}
\begin{centering}
\includegraphics[width=1.0\textwidth]{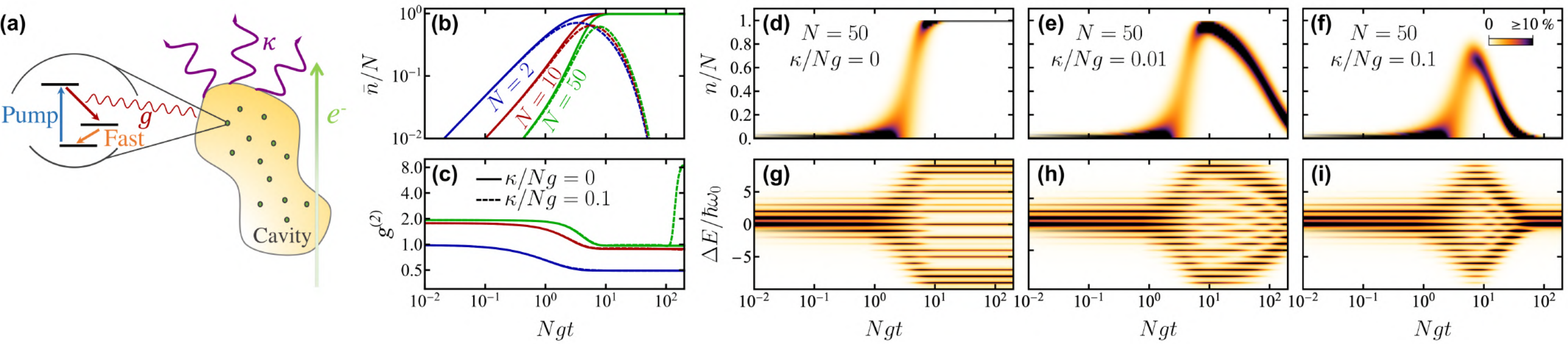}
\par\end{centering}
\caption{Interaction with an optical cavity coupled to pumped quantum emitters (QEs). (a) We consider a bosonic optical cavity (e.g., a Mie resonator) sustaining a single mode (frequency $\omega_0$, inelastic decay rate $\kappa$) and infiltrated with $N$ three-level QEs. Optical pumping prepares the emitters in their upper energy state at time $t=0$, from which they decay to an intermediate level by resonant coupling to the cavity mode at a rate $g$. (b,c) Temporal evolution of the average cavity mode population $\bar{n}$ (b) and second-order autocorrelation function at zero delay ${\rm g}^{(2)}$ (c) for $N=2$, 10, and 50 with $\kappa=0$ (solid curves) and $\kappa/Ng=0.1$ (dashed curves). (d-i) Evolution of the populations $p_n$ (d-f) and the electron spectra (g-i) as a function of the delay time for $N=50$ and different cavity decay rates: $\kappa/Ng=0$ (d,g), 0.01 (e,h) and 0.1 (f,i). We assume an electron-mode coupling $|\beta_0|=0.7$.}
\label{Fig3}
\end{figure*}

\section{Interaction with an optical cavity populated through pumped QEs}

As a feasible system to explore the above ideas, we consider an optical cavity (e.g., a Mie resonator, similar to those already probed by EELS \cite{HCR08}) hosting a spectrally isolated mode (frequency $\omega_0$, inelastic damping rate $\kappa$) fed by a number $N$ of 3-level QEs, as illustrated in Fig.\ \ref{Fig3}(a). A similar scheme applies to 4-level systems, which are also extensively used in experimental realizations of QEs coupled to optical cavities \cite{RR15}. The emitters are initialized in their excited state at $t=0$ (e.g., by optical pumping with a $\pi$ pulse), from which they decay to an intermediate state by coupling to the cavity at a rate $g$ (same for all QEs for simplicity) with a resonant transition frequency $\omega_0$. We further assume fast internal decay from the intermediate to the ground state, so that each QE interacts with the cavity only once. The combined probability $p_n^m$ for a cavity Fock state $|n\rangle$ with $m$ remaining excited emitters follows the equation of motion $dp_n^m/dt=g\left[n(m+1)p_{n-1}^{m+1}-(n+1)mp_n^m\right]+\kappa\left[(n+1)p_{n+1}^m-np_n^m\right]$, which we solve numerically to obtain the time-dependent distribution $p_n=\sum_{m=0}^N p_n^m$. Examples of the evolution of the resulting average mode population $\bar{n}$ and second-order autocorrelation ${\rm g}^{(2)}$ are shown in Fig.\ \ref{Fig3}(b,c). The latter starts at ${\rm g}^{(2)}=2(1-1/N)$ at $t=0$ and in the absence of damping evolves toward $1-1/N$ at long times (see Appendix\ \ref{populationnm}), as expected for an assembly of $N$ single-photon emitters. For finite cavity damping, $\bar{n}$ reaches a maximum $<N$, from which it exhibits an exponential decay, while ${\rm g}^{(2)}$ eventually jumps to large values when $\bar{n}$ becomes very small. For a sufficiently large number of QEs, we find a substantial average population while ${\rm g}^{(2)}$ varies from nearly 2 down to a quantum regimes characterized by $<1$. This evolution strongly affects the resulting electron spectra as a function of the time $t$ at which the electron-boson interaction occurs after pumping the QEs, shown in Fig.\ \ref{Fig3}(g-i) under the assumption that the interaction time is small compared with both $1/g$ and $1/\kappa$. The spectra initially resemble those of the thermal distributions of Fig.\ \ref{Fig2}, as expected from the ${\rm g}^{(2)}\approx2$ values, and gradually become similar to those of Fock states; for finite cavity damping, they undergo an attenuation similar to Fig.\ \ref{Fig2}(g) as $\sqrt{\bar{n}}|\beta_0|$ decreases [right part of Figs.\ \ref{Fig3}(h,i)]. To illustrate the feasibility of the assumed parameters, we consider a Mie resonance in a silicon sphere ($\epsilon=12$, radius $a$) for which we estimate (see Appendix\ \ref{Miecavity}) a sharp resonance of quality factor $\omega_0/\kappa\sim10^4$ for a size parameter $\omega_0a/c=2.6775$, which causes an increase in the decay rate of QEs embedded in it (Purcell effect \cite{P1946}) by an enhancement factor $g/g_0\sim100$, so for natural QE decay rates $g_0\sim$\,GHz and optical frequencies $\omega_0\sim100\,$THz, we have $\kappa/g\sim0.1$, while the coupling parameter for this mode is $|\beta_0|\sim0.03$ with 100-200\,keV electrons.

\section{Conclusion}

In summary, we have shown that the interaction of electron beams with near optical fields depends on both the quantum nature of the sample excitations (fermionic vs bosonic) and the statistics of their populations. For bosonic modes, the spectral distribution of losses and gains varies dramatically when comparing Fock, coherent, and thermal distributions. Our simulations reveal that these regimes can be explored by populating an optical cavity through optically pumped quantum emitters, for which we have elaborated a model based on realistic optical cavities (e.g., Mie resonators) infiltrated with gain atoms or molecules (e.g., rhodamine B). We predict that the autocorrelation functions of the population distributions are directly retrievable from the peak intensities in the electron spectra, thus opening an unexplored avenue for the study of the ultrafast plasmon, hot-electron, and phonon dynamics in optically pumped nanostructures using time-resolved electron microscope spectroscopy.

\section*{APPENDIX}

\appendix

\section{Interaction between fast electrons and optical modes: A relativistic quantum-optics approach}
\label{interaction}

The interaction of a beam electron with the evanescent field surrounding an illuminated nanostructure has been extensively studied under the assumption of a classical light field \cite{paper151,PLZ10,PZ12,paper272,paper311}. Here, we present a self-contained derivation of the theory needed to describe the interaction with nonclassical evanescent fields. But first, we start by revisiting the interaction with a classical field in a fully relativistic approach \cite{PZ12}. We represent the optical field by the scalar and vector potentials $\varphi$ and $\Ab$, and describe the interaction with the electron through the Dirac equation \cite{M1966,S1994}
\begin{align}
\left[\me c^2\beta+c\vec{\alpha}\cdot(\pb+\frac{e}{c}\Ab)-e\varphi\right]\Psi=\ii\hbar\frac{\partial\Psi}{\partial t},
\label{Dirac}
\end{align}
where $\pb=-\ii\hbar\nabla$ is the momentum operator and
\begin{align}
\beta=\left[\begin{array}{cc}
\mathcal{I} & 0 \\
0 & -\mathcal{I}
\end{array} \right], \quad\quad
\vec{\alpha}=\left[\begin{array}{cc}
0 & \vec{\sigma} \\
\vec{\sigma} & 0
\end{array} \right]
\nonumber
\end{align}
are Dirac matrices defined in terms of the $2\times2$ identity and Pauli matrices $\mathcal{I}$ and $\vec{\sigma}$. We now expand the electron 4-component spinor in terms of positive-energy plane waves (i.e., we neglect electron-positron pair creation processes \cite{S1994}) of well-defined relativistic momentum $\hbar\kb$ and energy $E_k=c\sqrt{\me^2c^2+\hbar^2k^2}$:
\begin{align}
\Psi=V^{-1/2}\sum_\kb \psi_\kb\;\ee^{\ii\kb\cdot\rb-\ii E_kt/\hbar}\;\Psi_\kb,
\label{expand}
\end{align}
where
\begin{align}
\Psi_\kb=\left[\begin{array}{c}
A_k\;\hat{\bf s} \\
B_k\;\vec{\sigma}\cdot\kb\;\hat{\bf s}
\end{array} \right]
\label{psikamp}
\end{align}
are the plane-wave spinor amplitudes, $\hat{\bf s}$ is a two-component spin unit vector, $V$ is the normalization volume, $A_k=\sqrt{(E_k+\me c^2)/2E_k}$, and $B_k=\hbar c/\sqrt{2E_k(E_k+\me c^2)}$. These plane waves are solutions of the noninteracting Dirac equation 
\begin{align}
\left(\me c^2\beta+c\vec{\alpha}\cdot\pb\right)\Psi_\kb=E_k\Psi_\kb,
\label{Dirac0}
\end{align}
so when inserting Eq.\ (\ref{expand}) into Eq.\ (\ref{Dirac}), we can replace $\me c^2\beta+c\vec{\alpha}\cdot\pb$ by $E_k$ inside the $\kb$ sum.

We focus in this work on incident electrons prepared in a state involving wave vectors $\kb$ tightly packed around a central value $\kb_0$, and further adopt the nonrecoil approximation by assuming that the interaction with the optical field effectively produces $\kb$ components also satisfying $|\kb-\kb_0|\ll k_0$. Under these conditions, we approximate the energy in Eq.\ (\ref{Dirac0}) by the first-order Taylor expansion $E_k\approx E_0+(\hbar^2c^2/E_0)\kb_0\cdot(\kb-\kb_0)$, where $E_0=E_{k_0}$. We further notice that $\kb$ inside the $\kb$ sum can be substituted by $-\ii\nabla$ outside it. Putting these elements together, we can recast Eq.\ (\ref{Dirac}) as
\begin{align}
\left[E_0-(\hbar^2c^2/E_0)\,\kb_0\cdot(\ii\nabla+\kb_0)+e\vec{\alpha}\cdot\Ab-e\varphi\right]\Psi=\ii\hbar\frac{\partial\Psi}{\partial t}.
\label{Dirac1}
\end{align}
We now approximate $\Psi_\kb\approx\Psi_{\kb_0}$ in Eq.\ (\ref{expand}) because only a small error is made when replacing $\kb$ by $\kb_0$ in the plane wave spinor amplitude [Eq.\ (\ref{psikamp})], so we have
\begin{align}
\Psi\approx\psi(\rb,t)\;\Psi_{\kb_0},
\label{psifinal}
\end{align}
where $\psi(\rb,t)=V^{-1/2}\sum_\kb \psi_\kb\;\ee^{\ii\kb\cdot\rb-\ii E_kt/\hbar}$ is a scalar electron wave function. Inserting Eq.\ (\ref{psifinal}) into Eq.\ (\ref{Dirac1}), multiplying both sides of the resulting equation by $\Psi_{\kb_0}^\dagger$ from the left, and noticing the relations $\Psi_\kb^\dagger\Psi_\kb=1$ and $\Psi_\kb^\dagger\vec{\alpha}\Psi_\kb=(\hbar c/E_k)\,\kb$ satisfied by the spinor plane waves, Eq.\ (\ref{Dirac1}) leads to the scalar equation
\begin{align}
\big[&E_0-(\hbar^2c^2/E_0)\,\kb_0\cdot(\ii\nabla+\kb_0)
\nonumber\\
&+(\hbar e c/E_0)\,\kb_0\cdot\Ab-e\varphi\big]\psi(\rb,t)=\ii\hbar\frac{\partial\psi(\rb,t)}{\partial t},
\nonumber
\end{align}
which we can recast using the electron velocity vector $\vb=(\hbar\kb_0/\me)/(1+E_0/\me c^2)$ and the ensuing relations $\hbar\kb_0=\me\vb\gamma$ and $E_0=\me c^2\gamma$ with $\gamma=1/\sqrt{1-v^2/c^2}$ as
\begin{align}
&\left[E_0-\hbar\vb\cdot(\ii\nabla+\kb_0)+(e\vb/c)\cdot\Ab-e\varphi\right]\psi(\rb,t)
\nonumber\\
&=\ii\hbar\frac{\partial\psi(\rb,t)}{\partial t}.
\label{Schret}
\end{align}
This expression, which results from the assumption of the nonrecoil approximation, constitutes an effective Schr\"odinger equation for the scalar amplitude $\psi(\rb,t)$ of the electron spinor [see Eq.\ (\ref{psifinal})], but we remark that it fully incorporates relativistic kinematics (i.e., $E_0$ and $\hbar\kb_0$ are the relativistic electron energy and momentum). Incidentally, the nonrecoil approximation allows us to assume that the spinor $\Psi_{\kb_0}$ is conserved during the interaction [see Eq.\ (\ref{psifinal})], and therefore, the electron spin $\hat{\bf s}$ does not change.

We now extend Eq.\ (\ref{Schret}) to account for nonclassical fields by working in a gauge in which $\varphi=0$ and substituting the vector potential by the operator \cite{M94} \[\hat{\Ab}=\sum_j(-\ii c/\omega_j)\left[\vec{\mathcal{E}}_j(\rb)\hat {a}_j-\vec{\mathcal{E}}_j^*(\rb)\hat {a}^\dagger_j\right],\] where the sum is extended over electromagnetic boson modes $j$ of creation and annihilation operators $\hat{a}^\dagger_j$ and $\hat{a}_j$, frequency $\omega_j$, and associated electric field $\vec{\mathcal{E}}_j(\rb)$. The wave function of the entire electron-field system $|\psi(\rb,t)\rangle=\sum_{\{n_j\}}\psi_{\{n_j\}}(\rb,t)|\{n_j\}\rangle$ is thus expanded to describe a distinct scalar electron wave function $\psi_{\{n_j\}}(\rb,t)$ for each of the possible number states $|\{n_j\}\rangle$ of the boson ensemble, so that we finally write the Schr\"odinger equation
\begin{align}
(\hat{\mathcal{H}}_0+\hat{\mathcal{H}}_1)|\psi(\rb,t)\rangle=\ii\hbar\frac{\partial|\psi(\rb,t)\rangle}{\partial t}
\label{Hfinal1}
\end{align}
with
\begin{align}
\hat{\mathcal{H}}_0&=\sum_j \hbar\omega_j\,\hat{a}^\dagger_j\hat{a}_j+E_0-\hbar\vb\cdot(\ii\nabla+\kb_0), \label{Hfinal2}\\
\hat{\mathcal{H}}_1&=(e\vb/c)\cdot\hat{\Ab},
\label{Hfinal3}
\end{align}
where the first term in $\hat{\mathcal{H}}_0$ is introduced to account for the noninteracting optical modes. We adopt these expressions in the main text assuming a single dominant boson mode corresponding to the index value $j=0$.

\section{Analytical solution of Eq.\ (\ref{fln}) in the main text}
\label{fanalytical}

Equation\ (\ref{fln}) is obtained from Eqs.\ (\ref{Hfinal1})-(\ref{Hfinal3}) by assuming a single mode $j=0$ and expressing the wave function in terms of amplitudes $f_\ell^n$ through Eq.\ (\ref{psi}). Importantly, Eq.\ (\ref{fln}) is formally equivalent to the Schr\"odinger equation for the coupling of a quantum harmonic oscillator to a classical perturbation $g(t)$. In order to show this in a tutorial way, we review some textbook physics and consider simpler forms of the free and interaction Hamiltonians $\hat{\mathcal{H}}_0=\hbar\omega_0\hat{a}^\dagger\hat{a}$ and $\hat{\mathcal{H}}_1=g^*(t)\hat{a}^\dagger+g(t)\hat{a}$, and write the Schr\"odinger equation $(\ii\hbar\partial/\partial t)|\psi_{\rm S}\rangle=\left(\hat{\mathcal{H}}_0+\hat{\mathcal{H}}_1\right)|\psi_{\rm S}\rangle$ with the wave function $|\psi_{\rm S}\rangle=\sum_n \alpha_n \ee^{-\ii n\omega_0t}|n\rangle$ expanded in terms of number states $|n\rangle$. The time-dependent amplitudes $\alpha_n$ satisfy the equation
\begin{align}
\ii\hbar\frac{d\alpha_n}{dt}=\sqrt{n}\,g^*\,\ee^{\ii\omega_0t}\alpha_{n-1}+\sqrt{n+1}\,g\,\ee^{-\ii\omega_0t}\alpha_{n+1},
\nonumber
\end{align}
which leads to Eq.\ (\ref{fln}) by performing the substitutions $t\rightarrow z/v$, $g(t)\ee^{-\ii\omega_0t}
\rightarrow-\ii\hbar v\,u(z)$, and $\alpha_n\rightarrow f_\ell^n$. Considering that $\ell+n$ is conserved during propagation according to Eq.\ (\ref{fln}), we need to separately apply this equivalence to each $\ell+n$ channel, the value of which is determined by the incident electron state (here assumed to be $\ell=0$) and each number state of the boson mode.

In order to obtain an analytical solution, it is convenient to move to the interaction picture, where the wave function $|\psi_{\rm I}\rangle=\ee^{\ii\hat{\mathcal{H}}_0t/\hbar}|\psi_{\rm S}\rangle$ satisfies $(\ii\hbar\partial/\partial t)|\psi_{\rm I}\rangle=\hat{\mathcal{H}}_{1{\rm I}}|\psi_{\rm I}\rangle$ and $\hat{\mathcal{H}}_{1{\rm I}}=\ee^{\ii\hat{\mathcal{H}}_0t/\hbar}\hat{\mathcal{H}}_1\ee^{-\ii\hat{\mathcal{H}}_0t/\hbar}
=g^*(t)\ee^{\ii\omega_0t}\hat{a}^\dagger+g(t)\ee^{-\ii\omega_0t}\hat{a}$. In terms of number states, we have
\begin{align}
|\psi_{\rm I}(t)\rangle=\sum_n \alpha_n(t)|n\rangle.
\nonumber
\end{align}
The sought-after analytical solution is then obtained as \cite{CN1965,LKB1970}
\begin{align}
|\psi_{\rm I}(t)\rangle=\hat{S}(t,t_0)|\psi_{\rm I}(t_0)\rangle,
\nonumber
\end{align}
where
\begin{align}
\hat{S}=\ee^{\ii\chi}\ee^{\beta_0^*\hat{a}^\dagger-\beta_0\hat{a}}
\nonumber
\end{align}
is the time-evolution operator,
\begin{align}
\beta_0(t,t_0)=\frac{\ii}{\hbar}\int_{t_0}^t dt' g(t')\ee^{-\ii\omega_0t'}
\label{beta0here}
\end{align}
is an effective coupling coefficient, and we have introduced a generally ignored time-dependent phase
\begin{align}
\chi(t,t_0)=\frac{-1}{\hbar}\int_{t_0}^t dt' {\rm Re}\left\{\beta_0(t',t_0)g^*(t')\ee^{\ii\omega_0t'}\right\}.
\nonumber
\end{align}
Incidentally, this solution can be directly verified by taking the time derivative of $\hat{S}$, for which it is useful to apply the relation $[\ee^{\hat{x}},\hat{y}]=C\ee^{\hat{x}}$ together with the Stone-von Neumann theorem $\ee^{\hat{x}+\hat{y}}=\ee^{\hat{x}}\ee^{\hat{y}}\ee^{-C/2}$, which are valid when the commutator $[\hat{x},\hat{y}]=C$ is a c-number. The latter theorem also allows us to write $\hat{S}=\ee^{\ii\chi}\ee^{\beta_0^*\hat{a}^\dagger}\ee^{-\beta_0\hat{a}}\ee^{-|\beta_0|^2/2}$. By Taylor expanding the exponential operators in this expression, we finally obtain the analytical result \cite{CN1965,LKB1970}
\begin{align}
\langle n|\hat{S}|n_0\rangle=&\ee^{\ii\chi}\sqrt{n_0!n!}\;\ee^{-|\beta_0|^2/2}(-\beta_0)^{n_0-n} \label{nSn}\\
&\times\sum_{n'} \frac{\left(-|\beta_0|^2\right)^{n'}}{n'!(n_0-n+n')!(n-n')!},
\nonumber
\end{align}
where the sum is restricted to the range ${\rm max}\{0,n-n_0\}\le n'\le n$. Equation\ (\ref{nSn}) is then applied in the main text to $n_0=\ell+n$ (i.e., the conserved initial sum of $\ell$ and $n$). Finally, we extend to integration region from $t_0\rightarrow-\infty$ to $t\rightarrow\infty$ and then Eq.\ (\ref{beta0here}) reduces to Eq.\ (\ref{beta1}) in the main text.

\begin{figure*}
\centering
\includegraphics[scale=0.5]{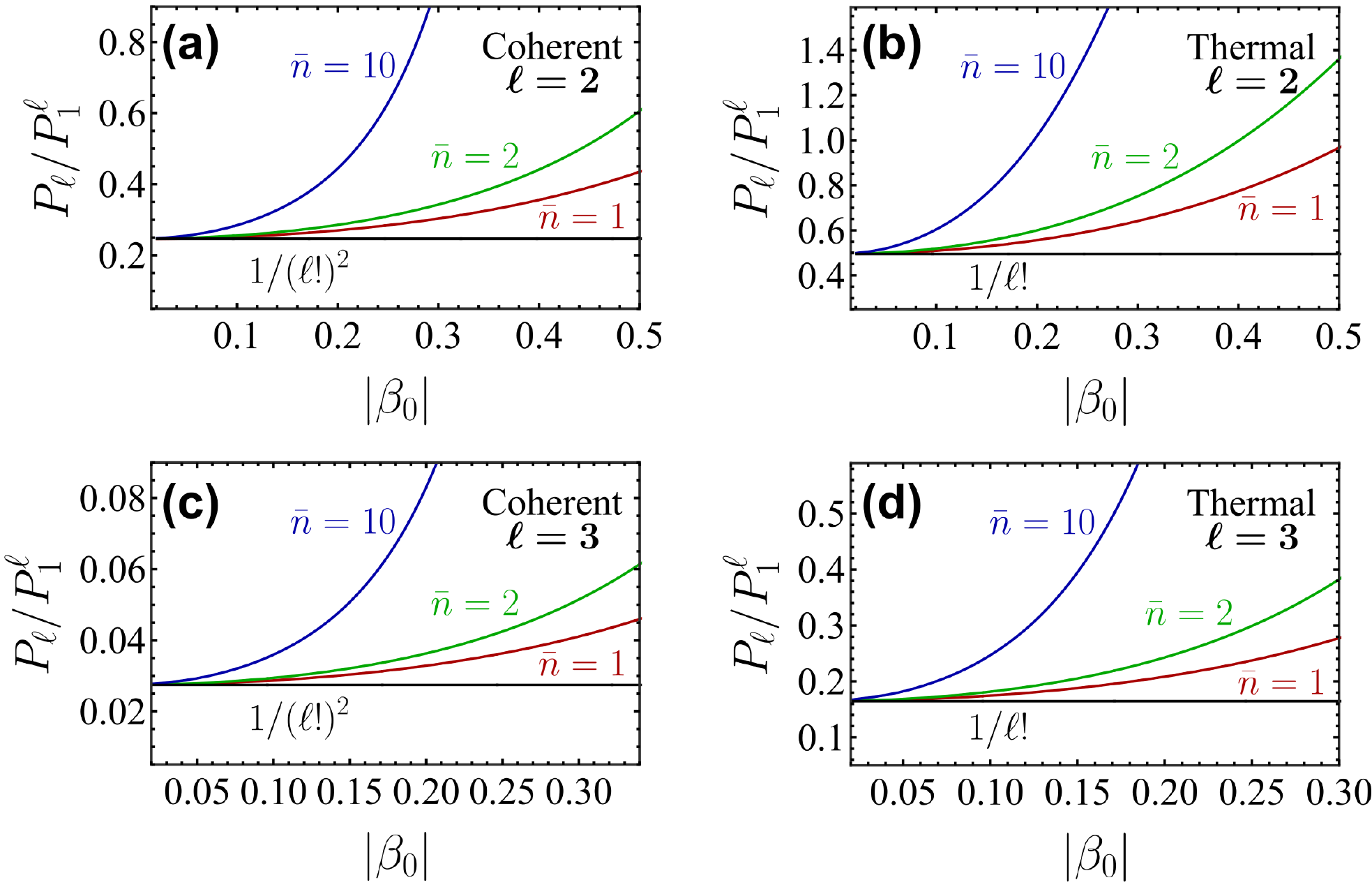}
\caption{(a,b) Ratio of the $\ell=2$ gain peak intensities for different average numbers of excitations (see color-coordinated labels) with coherent (a) and thermal (b) populations. (c,d) Same as (a,b) for $\ell=3$. Black horizontal lines corresond to the analytical weak-coupling limit.}
\label{FigS1}
\end{figure*}

\section{Electron-boson interaction in the weak-coupling limit}
\label{smallbeta}

A perturbative solution can be produced for Eq.\ (\ref{fln}) of the main text in the weak-coupling limit, provided the variations of all amplitudes $f_\ell^n$ are small during electron-boson interaction. When preparing the incident electron in $\ell=0$ and the boson in state $|n\rangle$, the nonvanishing elements of the perturbation series $f_\ell^n=\sum_{s=0}^\infty f_\ell^{n,s}$ satisfy the equations
\begin{align}
&\frac{df_{-1}^{n+1,1}}{dz}=\sqrt{n+1}\;u^*, \nonumber\\
&\frac{df_{1}^{n-1,1}}{dz}=-\sqrt{n}\;u, \nonumber\\
&\frac{df_{-2}^{n+2,2}}{dz}=\sqrt{n+2}\;u^*\,f_{-1}^{n+1,1}, \nonumber\\
&\frac{df_{0}^{n,2}}{dz}=\sqrt{n}\;u^*\,f_1^{n-1,1}-\sqrt{n+1}\;u\,f_{-1}^{n+1,1}, \nonumber\\
&\frac{df_{2}^{n-2,2}}{dz}=-\sqrt{n-1}\;u\,f_1^{n-1,1}, \nonumber\\
&\dots, \nonumber
\end{align}
where $u(z)=(e/\hbar\omega_0)\mathcal{E}_{0z}(z)\ee^{-\ii\omega_0z/v}$. After interaction, the first-order ($s=1$) amplitudes reduce to $f_{-1}^{n+1,1}(\infty)=\sqrt{n+1}\,\beta_0$ and $f_1^{n-1,1}(\infty)=-\sqrt{n}\,\beta_0$, which upon insertion into Eq.\ (\ref{Pl}) lead to 
\begin{subequations}
\begin{align}
&P_{-1}=(1+\bar{n})|\beta_0|^2, \\
&P_1=\bar{n}|\beta_0|^2,
\end{align}
\label{Pweak}
\end{subequations}  \noindent
where $\beta_0=\int dz\,u(z)$ [see Eq.\ (\ref{beta1}) in the main text] and $\bar{n}=\sum_{n=0}^\infty n\,p_n$ is the average population corresponding to the boson occupation distribution $p_n$. These expressions allow us to identify the weak-coupling limit condition as $\sqrt{\bar{n}}|\beta_0|\ll1$.

In the above series expansion, the lowest-order contribution to the $\ell>0$ gain peak corresponds to the amplitude $f_\ell^{n-\ell,\ell}$, which satisfies the equation
\begin{align}
\frac{df_{\ell}^{n-\ell,\ell}}{dz}=-\sqrt{n-\ell+1}\;u\,f_{\ell-1}^{n-\ell+1,\ell-1}. \nonumber
\end{align}
By iteratively solving this concatenated series of equations, we find the post-interaction solution
\begin{align}
&f_{\ell}^{n-\ell,\ell}(\infty)=(-1)^\ell \sqrt{n(n-1)\dots(n-\ell+1)}
\nonumber\\
&\times\int_{-\infty}^\infty dz_1\int_{-\infty}^{z_1}dz_2\cdots\int_{-\infty}^{z_{\ell-1}}dz_\ell
\;u(z_1)u(z_2)\cdots u(z_\ell)
\nonumber\\
&=\frac{(-\beta_0)^\ell}{\ell!}, \nonumber
\end{align}
where the rightmost expression is derived upon examination of the symmetry of the $\ell$-dimensional integrand upon permutation of its $\ell$ arguments \cite{AGD1965}, which allows us to push the upper integration limits to $\infty$ by creating $\ell!$ copies of it. The intensity of the $\ell>1$ gain peak thus becomes $P_\ell=|\beta_0|^{2\ell}\left\langle n(n-1)\cdots(n-\ell+1)\right\rangle/(\ell!)^2$, where $\langle\rangle$ denotes the average over the mode population. This leads to the powerful result
\begin{align}
\frac{P_\ell}{P_1^\ell}=\frac{{\rm g}^{(\ell)}}{(\ell!)^2}, \nonumber
\end{align}
where ${\rm g}^{(\ell)}=\left\langle n(n-1)\cdots(n-\ell+1)\right\rangle/\bar{n}^\ell$ is the $\ell^{\rm th}$-order correlation function at zero delay, which can then be directly inferred from a ratio of peak intensities measured in the transmitted electron spectrum. We present some illustrative examples in Fig.\ \ref{FigS1}.

For coherent states (i.e., a Poissonian distribution), we have ${\rm g}^{(\ell)}=1$ (with $\ell>0$) \cite{L1983}, leading to gain peak intensity ratios $P_\ell/P_1^\ell=(1/\ell!)^2$. In contrast, for a thermal distribution one has ${\rm g}^{(\ell)}=\ell!$ \cite{L1983}, which produces more intense gain peaks with $P_\ell/P_1^\ell=1/\ell!$ We stress that these results are valid only for weak interactions, as we are assuming that $|\beta|=\sqrt{\bar{n}}|\beta_0|\ll1$.

Incidentally, if we also assume $\bar{n}\gg1$ then $P_\ell$ is given by analytical expressions for coherent and thermal mode populations [see Eq.\ (f) in the main text and Sec.\ \ref{largen} below], for which, using the small argument approximation $\approx(z/2)^\ell/\ell!$ of both $J_\ell(z)$ and $I_\ell(z)$ with $\ell\ge0$ \cite{AS1972}, we find $P_\ell\approx|\beta|^{2\ell}/(\ell!)^2$ and $P_\ell\approx|\beta|^{2\ell}/\ell!$, respectively, therefore directly recovering the above results for the $P_\ell/P_1^\ell$ ratio. Additionally, we note that Fock states lead to the same ratio as coherent states in the $\bar{n}\gg1$ limit because they are characterized by ${\rm g}^{(\ell)}=\bar{n}(\bar{n}-1)\dots(\bar{n}-\ell+1)/\bar{n}^\ell\approx1$ (with $\ell>0$).

\section{Electron-boson interaction in the $\bar{n}\gg1$ limit}
\label{largen}

In the $\bar{n}\gg1$ limit, the bulk of the electron-boson interaction involves high $n$'s, which we consider to be much larger than the net number of quanta exchanges $\ell$. This condition is satisfied if the interaction-strength parameter is small ($|\beta_0|\ll1$), which is still compatible with a high total effective interaction $\bar{n}|\beta_0|^2\sim1$ for sufficiently large $\bar{n}$. We can then approximate both $\sqrt{n}$ and $\sqrt{n+1}$ by $\sqrt{n+\ell}$ in Eq.\ (\ref{fln}) of the main text; for each value of $n+\ell$, which is conserved during propagation of the electron amplitudes $f_\ell^n$ along $z$, the resulting equation coincides with Eq.\ (\ref{Pl}) of Ref.\ \onlinecite{paper272} for the PINEM interaction with an optical field $\mathcal{E}_z=\mathcal{E}_{0z}\sqrt{n+\ell}$, and therefore, we take from that reference the solution $f_\ell^n(\infty)=\sqrt{p_{n+\ell}}\ee^{\ii\ell\arg\{-\beta_0\}}J_\ell(2\sqrt{n+\ell}|\beta_0|)$, where $\beta_0$ is the electron-mode interaction parameter defined by Eq.\ (\ref{beta1}) in the main text and $p_{n+l}$ is the population distribution of the mode Fock state $|n+\ell\rangle$ before interaction with the electron. Because $n\gg|\ell|$, we can further approximate $n+\ell\approx n$ and write the probability associated with a net number $\ell$ of quanta exchanges [Eq.\ (\ref{Pl}) in the main text] as
\begin{align}
P_\ell\approx\sum_{n=0}^\infty p_nJ_\ell^2(2\sqrt{n}|\beta_0|).
\label{Psum}
\end{align}
For a boson prepared in the Fock state $|n\rangle$, the interaction probabilities reduce to $P_\ell=J_\ell^2(2|\beta|)$, where $\beta=\sqrt{\bar{n}}\beta_0$ and obviously the average mode population is $\bar{n}=n$. Likewise, for a coherent state the population distribution approaches a Gaussian $p_n\approx\ee^{-(n-\bar{n})^2/2\bar{n}}/\sqrt{2\pi\bar{n}}$ in the $\bar{n}\gg1$ limit \cite{F1968}, the width of which ($\sim\sqrt{\bar{n}}$) becomes increasingly small compared with the average population $\bar{n}$ as this one increases; we can thus approximate $n\approx\bar{n}$ inside the Bessel function of Eq.\ (\ref{Psum}), which leads to $P_\ell\approx J_\ell^2(2|\beta|)$, that is, the same result as for the Fock state. We thus conclude that in the large average population limit both Fock and coherent states of the boson mode produce the same types of electron spectra as observed in PINEM experiments.

The situation is however different for a chaotic thermal distribution $p_n=(1-\ee^{-\theta})\,\ee^{-n\theta}$ with average population $\bar{n}=1/(\ee^\theta-1)$, where $\theta=\hbar\omega_0/\kB T$ and $T$ is the mode temperature. We approach the $\bar{n}\gg1$ limit at high temperatures, for which $\theta\ll1$, and consequently, $\theta\approx1/\bar{n}$ and $p_n\approx\ee^{-n/\bar{n}}/\bar{n}$. Inserting these expressions into Eq.\ (\ref{Psum}) and approximating the sum as an integral with the change of variable $n/{\bar{n}}=x^2$, we find
\begin{align}
P_\ell\approx\int_0^\infty x\,dx\;\ee^{-x^2}J_\ell^2(2x|\beta|)=\ee^{-2|\beta|^2}I_\ell(2|\beta|^2),
\nonumber
\end{align}
where again $\beta=\sqrt{\bar{n}}\beta_0$, and the rightmost analytical equality (Eq.\ (6.633-2) of Ref.\ \cite{GR1980}) allows us to express the result in terms of the modified Bessel function $I_\ell$.

\section{Steady-state population of optically pumped bosonic and fermionic modes}
\label{weakcoupling}

\subsection{Fermionic mode}

A two-level system (states $j=0$,1 of energies $\hbar\varepsilon_j$) coupled to a monochromatic light field $\Eb(t)=\Eb_0\ee^{-\ii\omega t}+\Eb_0^*\ee^{\ii\omega t}$ constitutes a textbook example of light-matter interactions, commonly described through the optical Bloch equations \cite{SZ97}. The density matrix of the system satisfies the equation of motion
\begin{align}
\frac{d\hat{\rho}}{dt}=\frac{\ii}{\hbar}[\hat{\rho},\hat{\mathcal{H}}]+\frac{\gamma}{2}(2\hat{\sigma}\hat{\rho}\hat{\sigma}^\dagger-\hat{\sigma}^\dagger\hat{\sigma}\hat{\rho}-\hat{\rho}\hat{\sigma}^\dagger\hat{\sigma}),
\label{drhodt}
\end{align}
where the Hamiltonian $\hat{\mathcal{H}}=\sum_j\hbar\varepsilon_j|j\rangle\langle j|+g(t)\left(\hat{\sigma}^\dagger+\hat{\sigma}\right)$ incorporates the interaction with the transition dipole $p$ through the coupling energy $g(t)=-E(t)p$ (we assume $E_0p$ to be real and the field aligned with the dipole), and we define $\hat{\sigma}=|0\rangle\langle1|$ and  $\hat{\sigma}^\dagger=|1\rangle\langle0|$. Additionally, we account for inelastic $1\rightarrow0$ transition losses at a rate $\kappa$ through a Lindbladian term in Eq. (\ref{drhodt}). Writing the density matrix in the interaction picture as $\hat{\rho}=\sum_{j,j'}\rho_{jj'}\ee^{\ii(\varepsilon_{j'}-\varepsilon_j)t}|j\rangle\langle j'|$, Eq.\ (\ref{drhodt}) reduces to the Bloch equations
\begin{align}
&\dot{\bar{n}}=(-2/\hbar){\rm Im}\{\rho_{10}\,g\ee^{-\ii\omega_0t}\}-\kappa \bar{n}, \nonumber\\
&\dot{\rho}_{10}=(-\ii/\hbar)(1-2\bar{n})\,g\ee^{\ii\omega_0t}-\kappa \rho_{10}/2 \nonumber
\end{align}
for the average population $\bar{n}=\rho_{11}\equiv p_1$ and the coherence $\rho_{10}$; the other two elements of the density matrix are given by $\rho_{00}\equiv p_1=1-\bar{n}$ and $\rho_{01}=\rho_{10}^*$. At resonance ($\omega=\omega_0$), adopting the rotating-wave approximation (RWA), we have $g\ee^{\pm\ii\omega_0t}\approx-E_0p$, leading to the steady-state solution ($\dot{\bar{n}}=\dot{\rho}_{10}=0$)
\begin{align}
&\bar{n}=\frac{1}{2}\frac{1}{1+I_{\rm s}/I}, \label{nnFer}
\end{align}
which depends on the ratio of the light intensity $I=(c/2\pi)|E_0|^2$ to the saturation intensity of the system $I_{\rm s}=c(\hbar\kappa)^2/16\pi p^2$.

The interaction with a fast electron can be described following exactly the same formalism discussed in the main text for a bosonic mode, but replacing the commutating bosonic operator $\hat{a}$ by the anticommutating fermionic operator $\hat{\sigma}$. This prescription leads to Eq.\ (\ref{fln}) with $f_\ell^n$ vanishing unless $n=0$ or 1. In the weak electron-fermion coupling regime, this results in loss and gain probabilities
\begin{align}
&P_{-1}=p_0|\beta_0|^2=(1-\bar{n})|\beta_0|^2, \nonumber\\
&P_1=p_1|\beta_0|^2=\bar{n}|\beta_0|^2, \nonumber
\end{align}
where $|\beta_0|^2$ accounts for the electron-mode coupling strength [see Eq.\ (\ref{beta1})] and $\bar{n}$ is given by Eq. (\ref{nnFer}).

\subsection{Bosonic mode}
\label{weakboson}

A bosonic mode is described by an expression identical to Eq.\ (\ref{drhodt}) with $\hat{\sigma}$ and $\hat{\sigma}^\dagger$ replaced by the annihilation and creation operators $a$ and $a^\dagger$, which now satisfy the commutation relation $[\hat{a},\hat{a}^\dagger]=1$. Additionally, $j$ must be replaced by a boson occupation index $n=0,1,\dots$ that labels Fock states $|n\rangle$, while $\varepsilon_j$ must be substituted by the ladder frequencies $(n+1/2)\omega_0$. The boson density matrix admits a rigorous analytical solution \cite{CN1965,paper228}: $\hat{\rho}=|\xi(t)\rangle\langle\xi(t)|$, where $|\xi\rangle=\ee^{-|\xi|^2/2}\sum_{n=0}^\infty\left(\xi^n/\sqrt{n!}\right)\ee^{-\ii n\omega_0t}|n\rangle$ is a coherent state of amplitude $\xi$ (in the interaction picture) satisfying $a|\xi\rangle=\ee^{-\ii\omega_0t}\xi|\xi\rangle$ \cite{G1963} and describing a Poisonian distribution with occupation probabilities $p_n=\langle n|\hat{\rho}|n\rangle=\ee^{-|\xi|^2}|\xi|^{2n}/n!$ and average population $\bar{n}=|\xi|^2$. By inserting this expression of $\hat{\rho}$ into the equation of motion, one finds the solution $\xi(t)=(-\ii/\hbar)\int_{-\infty}^t dt' g(t')\ee^{\ii\omega_0t'-(t-t')\kappa/2}$ for the mode amplitude driven by and external classical perturbation $g(t)$. In particular, for the monochromatic light field considered above, this integral leads to $\xi(t)=(E_0p/\hbar)\big[\ee^{\ii(\omega_0-\omega)t}/(\omega_0-\omega-\ii\kappa/2)+\ee^{\ii(\omega_0+\omega)t}/(\omega_0+\omega-\ii\kappa/2)\big]$. Assuming again resonant illumination ($\omega=\omega_0$) and neglecting the off-resonance term $\propto\ee^{2\ii\omega_0t}$, the average population reduces to $\bar{n}=|\xi|^2\approx(2E_0p/\hbar\kappa)^2=I/2I_{\rm s}$. The interaction with an electron in the weak coupling regime is then given by Eqs.\ (\ref{Pweak}) with the mode occupation written as $\bar{n}=I/2I_{\rm s}$.

\section{Interaction with a dipolar mode and ensuing EELS probability}
\label{dipolarEELS}

We consider a dipolar mode of frequency $\omega_0$ characterized by a transition electric dipole moment $\pb$ placed at the origin. As an ansatz, we write the single-mode electric field as the one produced by this dipole, \[\vec{\mathcal{E}}_0=[k_0^2\pb+(\pb\cdot\nabla)\nabla]\ee^{\ii k_0r}/r,\] where $k_0=\omega_0/c$. The interaction parameter $\beta_0$ can be now calculated upon insertion of this field into Eq.\ (\ref{beta1}) in the main text. Integrating by parts, we find that each of the $z$ derivatives in the expression for $\vec{\mathcal{E}}_0$ can be replaced by $\ii\omega_0/v$. Finally, assuming a distance $b$ from the electron beam to the dipole and using the integral $\int dz\,\ee^{\ii k_0r-\ii\omega_0z/v}/r=2K_0(\zeta)$, where $r=\sqrt{b^2+z^2}$, $\zeta=\omega_0b/v\gamma$, and $\gamma=1/\sqrt{1-v^2/c^2}$ (see Eq.\ (3.914-4) in Ref. \cite{GR1980}, which we use here under the assumption that $k_0$ has an infinitesimal positive imaginary part), we readily find the expression \[\beta_0=(-2e\omega_0/\hbar v^2\gamma)\left[\ii p_x K_1(\zeta)+(p_z/\gamma)K_0(\zeta)\right],\] which is used in the main text.

In the weak-coupling regime (see Sec.\ \ref{weakcoupling}), the integral of the EELS probability over the mode spectral peak when the mode is initially depleted ($\bar{n}=0$) reduces to $P_{-1}=|\beta_0|^2=(2e\omega_0/\hbar v^2\gamma)^2\left[|p_x|^2 K_1^2(\zeta)+(|p_z|/\gamma)^2K_0^2(\zeta)\right]$. For an isotropic particle characterized by a triply-degenerate mode of electric dipoles along the Cartesian directions, the probability is given by the sum over the three polarization directions, which amounts to setting $p_x=p_y=p_z=p$; this leads to
\begin{align}
P_{-1}^{\rm isotropic}=|\beta_0|^2=|2e\omega_0p/\hbar v^2\gamma|^2f(\omega_0b/v\gamma),
\label{P1iso}
\end{align}
where $f(\zeta)=K_1^2(\zeta)+K_0^2(\zeta)/\gamma^2$.

In order to corroborate the correctness of the normalization of $\vec{\mathcal{E}}_0$ in the above ansatz, we compare $P_{-1}^{\rm isotropic}$ with the result derived from the classical EELS probability for an isotropic dipolar particle \cite{paper149}, $\Gamma_{\rm EELS,dip}(\omega)=(1/\hbar\pi)(2e\omega/v^2\gamma)^2f(\omega b/v\gamma)\,{\rm Im}\{\alpha(\omega)\}$, where $\alpha(\omega)$ is the polarizability. Linear response theory allows us to write the latter as \cite{PN1966} $\alpha(\omega)=(|p|^2/\hbar)\left[1/(\omega_0-\omega-\ii0^+)+1/(\omega_0+\omega+\ii0^+)\right]$ in terms of the mode dipole $p$ and frequency $\omega_0$, which upon insertion into the spectral integral $P_{-1}=\int_0^\infty d\omega \;\Gamma_{\rm EELS,dip}(\omega)$ reproduces Eq.\ (\ref{P1iso}), therefore confirming the ansatz.

\section{Electron-boson coupling strength for plasmonic cavities}
\label{plasmoncavity}

Plasmons in metallic nanoparticles constitute excellent candidates to explore the interaction between electrons and optical cavities. Here, we estimate the coupling parameter $|\beta_0|$ [see leading factor in Eq.\ (\ref{P1iso})] for two types of metallic nanoparticles in which the aspect ratio allows one to tune their frequency, also affecting the values of $|\beta_0|$; in particular, we consider prolate ellipsoids and spherical shells made of silver (permittivity $\epsilon(\omega)\approx\epsilonb-\wp^2/\omega(\omega+\ii/\tau)$ with $\epsilonb\approx4$, $\hbar\wp=9.17\,$eV, and $\hbar/\tau=21\,$meV \cite{JC1972}). Following similar methods as those of Ref.\ \cite{paper331}, a prolate ellipsoid of volume $V$ is found to exhibit a normal-to-the-symmetry-axis resonance frequency $\omega_0=\wp/\sqrt{\epsilonb-\epsilon_0}$ and an effective transition dipole $p\approx(1-\epsilon_0)\sqrt{\hbar\omega_0V/8\pi(\epsilonb-\epsilon_0)}$, where $\epsilon_0=1-1/L$, $L=(r^2/2)\Delta^{-3}\left[\pi/2-\arctan(1/\Delta)-\Delta/r^2\right]$ is the depolarization factor for a height-to-diameter aspect ratio $r>1$, and $\Delta=\sqrt{r^2-1}$. For a spherical metal shell of small thickness $t$ compared with the radius $a$, filled with a core dielectric $\epsilon_{\rm c}$, and treated in the $t\ll a$ limit, we find $\omega_0\approx\left(\wp/\sqrt{\epsilon_{\rm c}+2}\right)\sqrt{2t/a}$ and $p\approx\sqrt{3\hbar\omega_0a^3/2(\epsilon_{\rm c}+2)}$. Numerical inspection of these two types particles yields optimum coupling values that can reach $|\beta_0|\sim0.1$ with plasmon energies in the 1\,eV region, particle diameters of $\sim20\,$nm, and electron energies $\sim10\,$keV when considering either disk-like prolate ellipsoids (aspect ratio $r\sim5$) or thin shells ($t/a\sim0.2$) filled with silica ($\epsilon_{\rm c}=2$). The values adopted in the main paper are commensurate with these parameters.

\begin{figure*}
\centering
\includegraphics[scale=0.5]{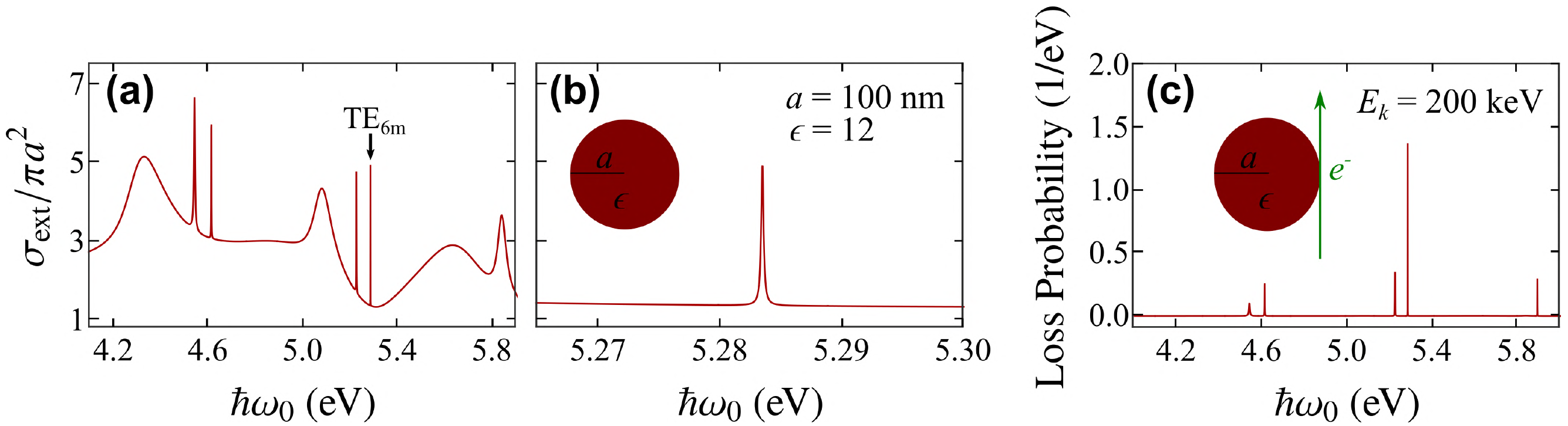}
\caption{(a,b) Optical extinction cross-section normalized to projected area for a silicon sphere ($\epsilon=12$) of radius $a=100\,$nm. Panel (b) shows a zoom around a TE mode with orbital angular number $l=6$ [see arrow in (a)]. (c) EELS probability for a 200\,keV electron under grazing incidence with respect to the sphere in (a,b).}
\label{FigS2}
\end{figure*}

\section{Interaction strength of quantum emitters and beam electrons with dielectric optical cavities}
\label{Miecavity}

High-index dielectric nanostructures can trap light with small radiative leakage. For example, a Si sphere of radius $a$ modeled with a permittivity $\epsilon=12$ exhibits a narrow resonance at a size parameter $\rho_0=\omega_0a/c\approx2.6775$ with a quality factor (frequency divided by width) $Q=\omega_0/\kappa\sim10^4$. Analytical EELS calculations based on a previously published formula \cite{paper149} predict a peak-integrated excitation probability $\sim10^{-3}$ (i.e., $|\beta_0|\sim0.03$) for grazingly passing electrons of 100-200\,keV kinetic energy (see illustrative calculation in Fig.\ \ref{FigS2}). Besides this relatively high probability, the large $Q$ value of the Mie resonance under consideration produces a high Purcell enhancement in quantum emitters (QEs) when they are embedded inside the structure, implying nearly perfect QE-cavity coupling and negligible radiative losses. Indeed, following a quantum optics formalism for dispersionless and lossless dielectrics \cite{GL91}, we can express the electromagnetic Green tensor as $G(\rb,\rb',\omega)=\sum_j \fb_j(\rb)\otimes\fb_j^*(\rb')/(\omega^2-\omega_j^2+\ii0^+)$, where the sum extends over photon modes $j$ in the presence of the cavity, and the mode functions are orthonormalized according to $\int d^3\rb\; \fb_j(\rb)\cdot\fb_{j'}^*(\rb)\epsilon(\rb)=\delta_{jj'}$; we now argue that the modes contributing to the Mie resonance $\omega_0$ should have similar spatial profiles inside the cavity, so we approximate $G(\rb,\rb',\omega)\approx\fb_0(\rb)\otimes\fb_0^*(\rb')/(\omega(\omega+\ii\kappa)-\omega_0^2)$ by phenomenologically introducing the resonance width $\kappa$; the Purcell enhancement factor EF is then proportional to the local-density of optical states (LDOS) normalized to the vacuum value \cite{paper102}, which can be calculated from $G$ as EF$\,\approx(-6\pi c^3/\omega){\rm Im}\{\nn\cdot G(\rb,\rb,\omega)\cdot\nn\}$, with $\rb=\rb'$ corresponding to the position of the QE; for resonant coupling $\omega=\omega_0$, arguing from the normalization condition that $|\fb_0|^2\sim1/\epsilon V$, where $V=4\pi a^3/3$ is the cavity volume, we find EF$\,\sim9Q/2\epsilon\rho_0^3\sim250$ for the cavity under consideration. We now envision QEs with a natural decay rate $g_0\sim\,$GHz, whose coupling rate increases to $g={\rm EF}\times g_0\sim10^2$GHz; for an optical frequency in the $\omega_0\sim100\,$THz range, the cavity damping rate is $\kappa\sim\omega_0/Q\sim10$GHz, thus leading to small values of $\kappa/g\sim0.1$ similar to those assumed in the main text.

\begin{figure*}
\centering
\includegraphics[scale=0.6]{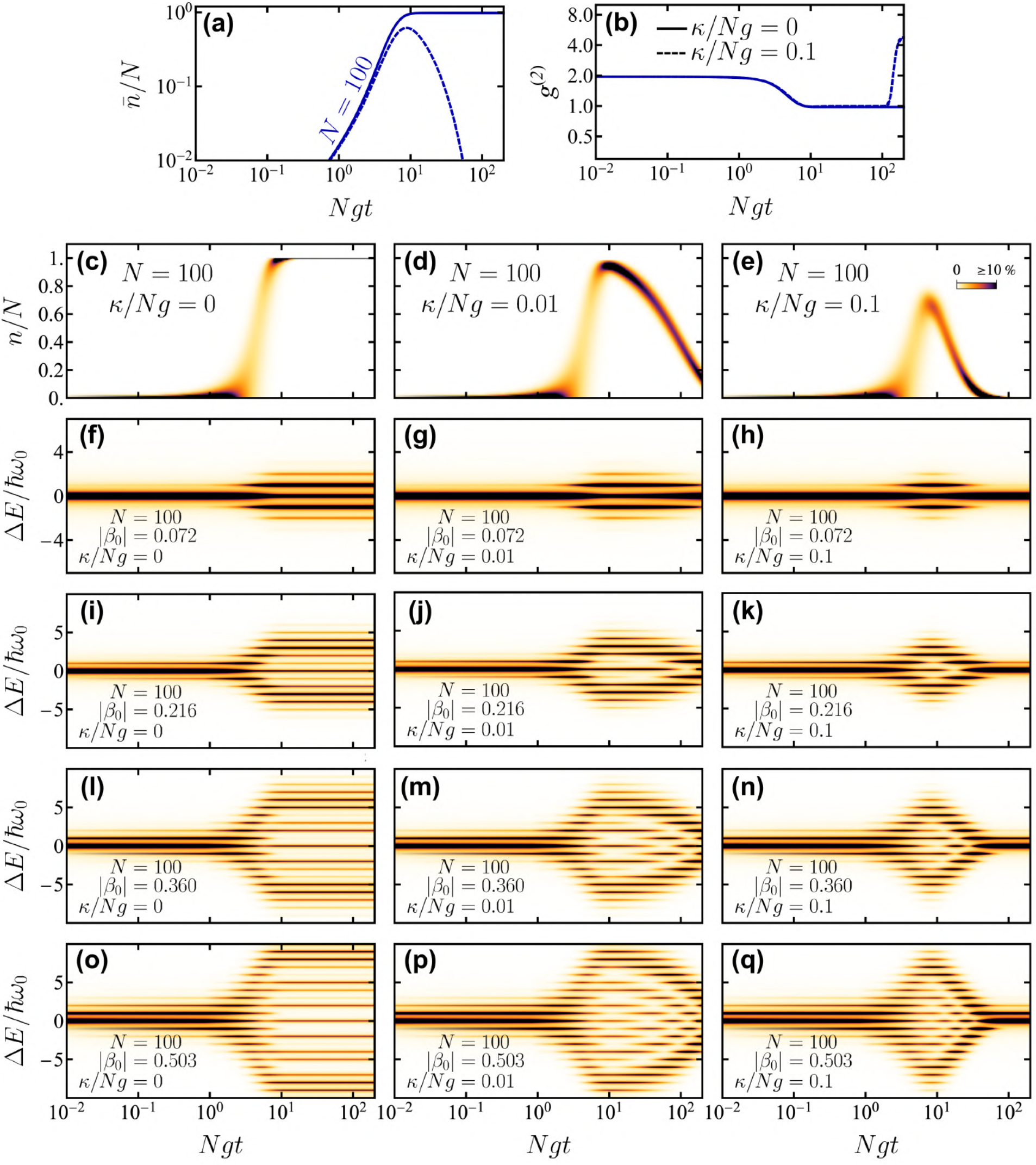}
\caption{(a-e) Same as Fig.\ 3(b-f) of the main text for $N=100$ QEs. (f-q) Evolution of electron spectra as a function of the delay time for $N=100$ and different values of the cavity decay rate $\kappa$ and electron-mode coupling $|\beta_0|$ (see labels).}
\label{FigS3}
\end{figure*}

\section{Population of a cavity coupled to an ensemble of quantum emitters}
\label{populationnm}

We consider an optical cavity hosting a bosonic excitation mode of frequency $\omega_0$ coupled to $N$ three-level QEs that are prepared in their excited state at time $t=0$ [see Fig.\ 3(a) in the main text]. The QEs can decay to their intermediate states by coupling to the cavity at a rate $g$, which we assume to be the same for all QEs. The transition frequency in this decay is taken to coincide with the cavity mode $\omega_0$. Once in their intermediate states, the QEs experience internal decay at a fast rate compared with $g$, so that they no longer interact with the cavity. The present analysis also applies to 4-level atoms such as those used in lasers with their intermediate-states transition matching $\omega_0$. We further incorporate an inelastic decay of the cavity mode at a rate $\kappa$, associated with either intrinsic losses in the materials or radiative emission.

We characterize the QEs-cavity system in terms of the time-dependent probabilities $p_n^m(t)$ for the cavity to be in the Fock state $|n\rangle$ while $m$ emitters are still in their excited states. The equation of motion for these probabilities can be readily written as
\begin{align}
\frac{dp_n^m}{dt}=&g\left[n(m+1)p_{n-1}^{m+1}-(n+1)mp_n^m\right]
\nonumber\\
&+\kappa\left[(n+1)p_{n+1}^m-np_n^m\right]
\label{pnm}
\end{align}
with the initial condition $p_n^m(0)=\delta_{n0}\delta_{mN}$. The $p_{n-1}^{m+1}$ term in the right-hand side of Eq.\ (\ref{pnm}) contributes to increase the $p_n^m$ probability due to the decay of an emitter [i.e., when going from $m+1$ to $m$, so it must be proportional to the number of excited QEs before the decay, $m+1$] that transfers its energy to the cavity [i.e., from $n-1$ to $n$, so it is proportional to $n$]; this type of process also subtracts probability from $p_n^m$, as described by the negative term proportional to $g$ in Eq.\ (\ref{pnm}); similar reasoning leads to the remaining two terms proportional to $\kappa$ to account for inelastic cavity losses. Upon inspection, we find that the total probability $\sum_{nm}p_n^m=1$ is conserved. Additionally, the total number of excitations in the system cannot exceed $N$, so $p_n^m=0$ if $n\ge N$ or $m\ge N$.

We are interested in the cavity occupation distribution $p_n=\sum_m p_n^m$, the average population $\bar{n}=\sum_n n\,p_n$, and how they influence the coupling to a passing electron. As a rough way of characterizing the population distribution, we consider the  instantaneous second-order correlation ${\rm g}^{(2)}=(\bar{n^2}-\bar{n})/\bar{n}^2$, where $\bar{n^2}=\sum_n n^2p_n$. At small times right after pumping the QEs to their excited states, ${\rm g}^{(2)}$ can be obtained analytically by Taylor expanding $p_n^m$; we find $\bar{n}=Ngt+O(t^2)$ and $\bar{n^2}-\bar{n}=2N(N-1)(gt)^2+O(t^3)$, from which ${\rm g}^{(2)}=2(1-1/N)$ at $t=0^+$. By numerically solving Eq.\ (\ref{pnm}), we observe that ${\rm g}^{(2)}$ is generally a decreasing function of $gt$ (see Fig.\ 2 in the main text). In particular, for $\kappa=0$ the number of excitations in the system must be conserved (i.e., $n+m=N$), and eventually all of them are transferred to the cavity, leading to the asymptotic solution $p_n^m(\infty)=\delta_{nN}\delta_{m0}$, which in turn produces ${\rm g}^{(2)}=1-1/N$. Incidentally, a large increase in ${\rm g}^{(2)}$ is eventually observed when $\bar{n}$ drops to very small values in lossy cavities.

We present additional results for cavities with $N=100$ QEs in Fig.\ \ref{FigS3}.

\acknowledgments

We thank Vahagn Mkhitaryan, Matteo Lostaglio, and Sophie Meuret for helpful and enjoyable discussions. This work has been supported in part by the Spanish MINECO (MAT2017-88492-R and SEV2015- 0522), ERC (Advanced Grant 789104-eNANO), the Catalan CERCA Program, and Fundaci\'{o} Privada Cellex. V.D.G. acknowledges support from the EU through a Marie Sk\l{}odowska-Curie grant (COFUND-DP, H2020-MSCA-COFUND-2014, GA n 665884).


\begin{thebibliography}{56}
\expandafter\ifx\csname natexlab\endcsname\relax\def\natexlab#1{#1}\fi
\expandafter\ifx\csname bibnamefont\endcsname\relax
  \def\bibnamefont#1{#1}\fi
\expandafter\ifx\csname bibfnamefont\endcsname\relax
  \def\bibfnamefont#1{#1}\fi
\expandafter\ifx\csname citenamefont\endcsname\relax
  \def\citenamefont#1{#1}\fi
\expandafter\ifx\csname url\endcsname\relax
  \def\url#1{\texttt{#1}}\fi
\expandafter\ifx\csname urlprefix\endcsname\relax\def\urlprefix{URL }\fi
\providecommand{\bibinfo}[2]{#2}
\providecommand{\eprint}[2][]{\url{#2}}

\bibitem[{\citenamefont{Howie}(2003)}]{H03}
\bibinfo{author}{\bibfnamefont{A.}~\bibnamefont{Howie}},
  \bibinfo{journal}{Micron} \textbf{\bibinfo{volume}{34}}, \bibinfo{pages}{121}
  (\bibinfo{year}{2003}).

\bibitem[{\citenamefont{Mkhoyan et~al.}(2007)\citenamefont{Mkhoyan, Babinec,
  Maccagnano, Kirkland, and Silcox}}]{MBM07}
\bibinfo{author}{\bibfnamefont{K.~A.} \bibnamefont{Mkhoyan}},
  \bibinfo{author}{\bibfnamefont{T.}~\bibnamefont{Babinec}},
  \bibinfo{author}{\bibfnamefont{S.~E.} \bibnamefont{Maccagnano}},
  \bibinfo{author}{\bibfnamefont{E.~J.} \bibnamefont{Kirkland}},
  \bibnamefont{and} \bibinfo{author}{\bibfnamefont{J.}~\bibnamefont{Silcox}},
  \bibinfo{journal}{Ultramicroscopy} \textbf{\bibinfo{volume}{107}},
  \bibinfo{pages}{345} (\bibinfo{year}{2007}).

\bibitem[{\citenamefont{Krivanek et~al.}(2014)\citenamefont{Krivanek, Lovejoy,
  Dellby, Aoki, Carpenter, Rez, Soignard, Zhu, Batson, Lagos et~al.}}]{KLD14}
\bibinfo{author}{\bibfnamefont{O.~L.} \bibnamefont{Krivanek}},
  \bibinfo{author}{\bibfnamefont{T.~C.} \bibnamefont{Lovejoy}},
  \bibinfo{author}{\bibfnamefont{N.}~\bibnamefont{Dellby}},
  \bibinfo{author}{\bibfnamefont{T.}~\bibnamefont{Aoki}},
  \bibinfo{author}{\bibfnamefont{R.~W.} \bibnamefont{Carpenter}},
  \bibinfo{author}{\bibfnamefont{P.}~\bibnamefont{Rez}},
  \bibinfo{author}{\bibfnamefont{E.}~\bibnamefont{Soignard}},
  \bibinfo{author}{\bibfnamefont{J.}~\bibnamefont{Zhu}},
  \bibinfo{author}{\bibfnamefont{P.~E.} \bibnamefont{Batson}},
  \bibinfo{author}{\bibfnamefont{M.~J.} \bibnamefont{Lagos}},
  \bibnamefont{et~al.}, \bibinfo{journal}{Nature}
  \textbf{\bibinfo{volume}{514}}, \bibinfo{pages}{209} (\bibinfo{year}{2014}).

\bibitem[{\citenamefont{Muller et~al.}(2008)\citenamefont{Muller, {Fitting
  Kourkoutis}, Murfitt, Song, Hwang, Silcox, Dellby, and Krivanek}}]{MKM08}
\bibinfo{author}{\bibfnamefont{D.~A.} \bibnamefont{Muller}},
  \bibinfo{author}{\bibfnamefont{L.}~\bibnamefont{{Fitting Kourkoutis}}},
  \bibinfo{author}{\bibfnamefont{M.}~\bibnamefont{Murfitt}},
  \bibinfo{author}{\bibfnamefont{J.~H.} \bibnamefont{Song}},
  \bibinfo{author}{\bibfnamefont{H.~Y.} \bibnamefont{Hwang}},
  \bibinfo{author}{\bibfnamefont{J.}~\bibnamefont{Silcox}},
  \bibinfo{author}{\bibfnamefont{N.}~\bibnamefont{Dellby}}, \bibnamefont{and}
  \bibinfo{author}{\bibfnamefont{O.~L.} \bibnamefont{Krivanek}},
  \bibinfo{journal}{Science} \textbf{\bibinfo{volume}{319}},
  \bibinfo{pages}{1073} (\bibinfo{year}{2008}).

\bibitem[{\citenamefont{Zhu et~al.}(2012)\citenamefont{Zhu, Radtke, and
  Botton}}]{ZRB12}
\bibinfo{author}{\bibfnamefont{G.}~\bibnamefont{Zhu}},
  \bibinfo{author}{\bibfnamefont{G.}~\bibnamefont{Radtke}}, \bibnamefont{and}
  \bibinfo{author}{\bibfnamefont{G.~A.} \bibnamefont{Botton}},
  \bibinfo{journal}{Nature} \textbf{\bibinfo{volume}{490}},
  \bibinfo{pages}{384} (\bibinfo{year}{2012}).

\bibitem[{\citenamefont{{Garc\'{\i}a de Abajo}}(2010)}]{paper149}
\bibinfo{author}{\bibfnamefont{F.~J.} \bibnamefont{{Garc\'{\i}a de Abajo}}},
  \bibinfo{journal}{Rev.\ Mod.\ Phys.} \textbf{\bibinfo{volume}{82}},
  \bibinfo{pages}{209} (\bibinfo{year}{2010}).

\bibitem[{\citenamefont{Rossouw and Botton}(2013)}]{RB13}
\bibinfo{author}{\bibfnamefont{D.}~\bibnamefont{Rossouw}} \bibnamefont{and}
  \bibinfo{author}{\bibfnamefont{G.~A.} \bibnamefont{Botton}},
  \bibinfo{journal}{Phys.\ Rev.\ Lett.} \textbf{\bibinfo{volume}{110}},
  \bibinfo{pages}{066801} (\bibinfo{year}{2013}).

\bibitem[{\citenamefont{Kociak and Stephan}(2014)}]{KS14}
\bibinfo{author}{\bibfnamefont{M.}~\bibnamefont{Kociak}} \bibnamefont{and}
  \bibinfo{author}{\bibfnamefont{O.}~\bibnamefont{Stephan}},
  \bibinfo{journal}{Chem.\ Soc.\ Rev.} \textbf{\bibinfo{volume}{43}},
  \bibinfo{pages}{3865} (\bibinfo{year}{2014}).

\bibitem[{\citenamefont{Anton~H\"orl and Hohenester}(2015)}]{HTH15}
\bibinfo{author}{\bibfnamefont{A.~T.} \bibnamefont{Anton~H\"orl}}
  \bibnamefont{and}
  \bibinfo{author}{\bibfnamefont{U.}~\bibnamefont{Hohenester}},
  \bibinfo{journal}{ACS\ Photon.} \textbf{\bibinfo{volume}{2}},
  \bibinfo{pages}{1429} (\bibinfo{year}{2015}).

\bibitem[{\citenamefont{Guzzinati et~al.}(2017)\citenamefont{Guzzinati, Beche,
  Lourenco-Martins, Martin, Kociak, and Verbeeck}}]{GBL17}
\bibinfo{author}{\bibfnamefont{G.}~\bibnamefont{Guzzinati}},
  \bibinfo{author}{\bibfnamefont{A.}~\bibnamefont{Beche}},
  \bibinfo{author}{\bibfnamefont{H.}~\bibnamefont{Lourenco-Martins}},
  \bibinfo{author}{\bibfnamefont{J.}~\bibnamefont{Martin}},
  \bibinfo{author}{\bibfnamefont{M.}~\bibnamefont{Kociak}}, \bibnamefont{and}
  \bibinfo{author}{\bibfnamefont{J.}~\bibnamefont{Verbeeck}},
  \bibinfo{journal}{Nat.\ Commun.} \textbf{\bibinfo{volume}{8}},
  \bibinfo{pages}{14999} (\bibinfo{year}{2017}).

\bibitem[{\citenamefont{Lagos et~al.}(2017)\citenamefont{Lagos, Tr\"ugler,
  Hohenester, and Batson}}]{LTHB17}
\bibinfo{author}{\bibfnamefont{M.~J.} \bibnamefont{Lagos}},
  \bibinfo{author}{\bibfnamefont{A.}~\bibnamefont{Tr\"ugler}},
  \bibinfo{author}{\bibfnamefont{U.}~\bibnamefont{Hohenester}},
  \bibnamefont{and} \bibinfo{author}{\bibfnamefont{P.~E.}
  \bibnamefont{Batson}}, \bibinfo{journal}{Nature}
  \textbf{\bibinfo{volume}{543}}, \bibinfo{pages}{529} (\bibinfo{year}{2017}).

\bibitem[{\citenamefont{Barwick et~al.}(2009)\citenamefont{Barwick, Flannigan,
  and Zewail}}]{BFZ09}
\bibinfo{author}{\bibfnamefont{B.}~\bibnamefont{Barwick}},
  \bibinfo{author}{\bibfnamefont{D.~J.} \bibnamefont{Flannigan}},
  \bibnamefont{and} \bibinfo{author}{\bibfnamefont{A.~H.}
  \bibnamefont{Zewail}}, \bibinfo{journal}{Nature}
  \textbf{\bibinfo{volume}{462}}, \bibinfo{pages}{902} (\bibinfo{year}{2009}).

\bibitem[{\citenamefont{Ryabov and Baum}(2016)}]{RB16}
\bibinfo{author}{\bibfnamefont{A.}~\bibnamefont{Ryabov}} \bibnamefont{and}
  \bibinfo{author}{\bibfnamefont{P.}~\bibnamefont{Baum}},
  \bibinfo{journal}{Science} \textbf{\bibinfo{volume}{353}},
  \bibinfo{pages}{374} (\bibinfo{year}{2016}).

\bibitem[{\citenamefont{Koz\'ak et~al.}(2017)\citenamefont{Koz\'ak, McNeur,
  Leedle, Deng, Sch\"onenberger, Ruehl, Hartl, Harris, Byer, and
  Hommelhoff}}]{KML17}
\bibinfo{author}{\bibfnamefont{M.}~\bibnamefont{Koz\'ak}},
  \bibinfo{author}{\bibfnamefont{J.}~\bibnamefont{McNeur}},
  \bibinfo{author}{\bibfnamefont{K.~J.} \bibnamefont{Leedle}},
  \bibinfo{author}{\bibfnamefont{H.}~\bibnamefont{Deng}},
  \bibinfo{author}{\bibfnamefont{N.}~\bibnamefont{Sch\"onenberger}},
  \bibinfo{author}{\bibfnamefont{A.}~\bibnamefont{Ruehl}},
  \bibinfo{author}{\bibfnamefont{I.}~\bibnamefont{Hartl}},
  \bibinfo{author}{\bibfnamefont{J.~S.} \bibnamefont{Harris}},
  \bibinfo{author}{\bibfnamefont{R.~L.} \bibnamefont{Byer}}, \bibnamefont{and}
  \bibinfo{author}{\bibfnamefont{P.}~\bibnamefont{Hommelhoff}},
  \bibinfo{journal}{Nat.\ Commun.} \textbf{\bibinfo{volume}{8}},
  \bibinfo{pages}{14342} (\bibinfo{year}{2017}).

\bibitem[{\citenamefont{{Garc\'{\i}a de Abajo}
  et~al.}(2010)\citenamefont{{Garc\'{\i}a de Abajo}, {Asenjo Garcia}, and
  Kociak}}]{paper151}
\bibinfo{author}{\bibfnamefont{F.~J.} \bibnamefont{{Garc\'{\i}a de Abajo}}},
  \bibinfo{author}{\bibfnamefont{A.}~\bibnamefont{{Asenjo Garcia}}},
  \bibnamefont{and} \bibinfo{author}{\bibfnamefont{M.}~\bibnamefont{Kociak}},
  \bibinfo{journal}{Nano\ Lett.} \textbf{\bibinfo{volume}{10}},
  \bibinfo{pages}{1859} (\bibinfo{year}{2010}).

\bibitem[{\citenamefont{Park et~al.}(2010)\citenamefont{Park, Lin, and
  Zewail}}]{PLZ10}
\bibinfo{author}{\bibfnamefont{S.~T.} \bibnamefont{Park}},
  \bibinfo{author}{\bibfnamefont{M.}~\bibnamefont{Lin}}, \bibnamefont{and}
  \bibinfo{author}{\bibfnamefont{A.~H.} \bibnamefont{Zewail}},
  \bibinfo{journal}{New\ J.\ Phys.} \textbf{\bibinfo{volume}{12}},
  \bibinfo{pages}{123028} (\bibinfo{year}{2010}).

\bibitem[{\citenamefont{Kirchner et~al.}(2014)\citenamefont{Kirchner, Gliserin,
  Krausz, and Baum}}]{KGK14}
\bibinfo{author}{\bibfnamefont{F.~O.} \bibnamefont{Kirchner}},
  \bibinfo{author}{\bibfnamefont{A.}~\bibnamefont{Gliserin}},
  \bibinfo{author}{\bibfnamefont{F.}~\bibnamefont{Krausz}}, \bibnamefont{and}
  \bibinfo{author}{\bibfnamefont{P.}~\bibnamefont{Baum}},
  \bibinfo{journal}{Nat.\ Photon.} \textbf{\bibinfo{volume}{8}},
  \bibinfo{pages}{52} (\bibinfo{year}{2014}).

\bibitem[{\citenamefont{Piazza et~al.}(2015)\citenamefont{Piazza, Lummen,
  {Qui\~{n}onez}, Murooka, Reed, Barwick, and Carbone}}]{PLQ15}
\bibinfo{author}{\bibfnamefont{L.}~\bibnamefont{Piazza}},
  \bibinfo{author}{\bibfnamefont{T.~T.~A.} \bibnamefont{Lummen}},
  \bibinfo{author}{\bibfnamefont{E.}~\bibnamefont{{Qui\~{n}onez}}},
  \bibinfo{author}{\bibfnamefont{Y.}~\bibnamefont{Murooka}},
  \bibinfo{author}{\bibfnamefont{B.}~\bibnamefont{Reed}},
  \bibinfo{author}{\bibfnamefont{B.}~\bibnamefont{Barwick}}, \bibnamefont{and}
  \bibinfo{author}{\bibfnamefont{F.}~\bibnamefont{Carbone}},
  \bibinfo{journal}{Nat.\ Commun.} \textbf{\bibinfo{volume}{6}},
  \bibinfo{pages}{6407} (\bibinfo{year}{2015}).

\bibitem[{\citenamefont{Feist et~al.}(2015)\citenamefont{Feist, Echternkamp,
  Schauss, Yalunin, Sch\"afer, and Ropers}}]{FES15}
\bibinfo{author}{\bibfnamefont{A.}~\bibnamefont{Feist}},
  \bibinfo{author}{\bibfnamefont{K.~E.} \bibnamefont{Echternkamp}},
  \bibinfo{author}{\bibfnamefont{J.}~\bibnamefont{Schauss}},
  \bibinfo{author}{\bibfnamefont{S.~V.} \bibnamefont{Yalunin}},
  \bibinfo{author}{\bibfnamefont{S.}~\bibnamefont{Sch\"afer}},
  \bibnamefont{and} \bibinfo{author}{\bibfnamefont{C.}~\bibnamefont{Ropers}},
  \bibinfo{journal}{Nature} \textbf{\bibinfo{volume}{521}},
  \bibinfo{pages}{200} (\bibinfo{year}{2015}).

\bibitem[{\citenamefont{Lummen et~al.}(2016)\citenamefont{Lummen, Lamb,
  Berruto, LaGrange, Negro, {Garc\'{\i}a de Abajo}, McGrouther, Barwick, and
  Carbone}}]{paper282}
\bibinfo{author}{\bibfnamefont{T.~T.~A.} \bibnamefont{Lummen}},
  \bibinfo{author}{\bibfnamefont{R.~J.} \bibnamefont{Lamb}},
  \bibinfo{author}{\bibfnamefont{G.}~\bibnamefont{Berruto}},
  \bibinfo{author}{\bibfnamefont{T.}~\bibnamefont{LaGrange}},
  \bibinfo{author}{\bibfnamefont{L.~D.} \bibnamefont{Negro}},
  \bibinfo{author}{\bibfnamefont{F.~J.} \bibnamefont{{Garc\'{\i}a de Abajo}}},
  \bibinfo{author}{\bibfnamefont{D.}~\bibnamefont{McGrouther}},
  \bibinfo{author}{\bibfnamefont{B.}~\bibnamefont{Barwick}}, \bibnamefont{and}
  \bibinfo{author}{\bibfnamefont{F.}~\bibnamefont{Carbone}},
  \bibinfo{journal}{Nat.\ Commun.} \textbf{\bibinfo{volume}{7}},
  \bibinfo{pages}{13156} (\bibinfo{year}{2016}).

\bibitem[{\citenamefont{Echternkamp et~al.}(2016)\citenamefont{Echternkamp,
  Feist, Sch\"{a}fer, and Ropers}}]{EFS16}
\bibinfo{author}{\bibfnamefont{K.~E.} \bibnamefont{Echternkamp}},
  \bibinfo{author}{\bibfnamefont{A.}~\bibnamefont{Feist}},
  \bibinfo{author}{\bibfnamefont{S.}~\bibnamefont{Sch\"{a}fer}},
  \bibnamefont{and} \bibinfo{author}{\bibfnamefont{C.}~\bibnamefont{Ropers}},
  \bibinfo{journal}{Nat.\ Phys.} \textbf{\bibinfo{volume}{12}},
  \bibinfo{pages}{1000} (\bibinfo{year}{2016}).

\bibitem[{\citenamefont{Feist et~al.}(2017)\citenamefont{Feist, Bach,
  N.~Rubiano~{da Silva}, M\"{a}ller, Priebe, Domr\"{a}se, Gatzmann, Rost,
  Schauss, Strauch et~al.}}]{FBR17}
\bibinfo{author}{\bibfnamefont{A.}~\bibnamefont{Feist}},
  \bibinfo{author}{\bibfnamefont{N.}~\bibnamefont{Bach}},
  \bibinfo{author}{\bibfnamefont{T.~D.} \bibnamefont{N.~Rubiano~{da Silva}}},
  \bibinfo{author}{\bibfnamefont{M.}~\bibnamefont{M\"{a}ller}},
  \bibinfo{author}{\bibfnamefont{K.~E.} \bibnamefont{Priebe}},
  \bibinfo{author}{\bibfnamefont{T.}~\bibnamefont{Domr\"{a}se}},
  \bibinfo{author}{\bibfnamefont{J.~G.} \bibnamefont{Gatzmann}},
  \bibinfo{author}{\bibfnamefont{S.}~\bibnamefont{Rost}},
  \bibinfo{author}{\bibfnamefont{J.}~\bibnamefont{Schauss}},
  \bibinfo{author}{\bibfnamefont{S.}~\bibnamefont{Strauch}},
  \bibnamefont{et~al.}, \bibinfo{journal}{Ultramicroscopy}
  \textbf{\bibinfo{volume}{176}}, \bibinfo{pages}{63} (\bibinfo{year}{2017}).

\bibitem[{\citenamefont{Priebe et~al.}(2017)\citenamefont{Priebe, Rathje,
  Yalunin, Hohage, Feist, Sch\"{a}fer, and Ropers}}]{PRY17}
\bibinfo{author}{\bibfnamefont{K.~E.} \bibnamefont{Priebe}},
  \bibinfo{author}{\bibfnamefont{C.}~\bibnamefont{Rathje}},
  \bibinfo{author}{\bibfnamefont{S.~V.} \bibnamefont{Yalunin}},
  \bibinfo{author}{\bibfnamefont{T.}~\bibnamefont{Hohage}},
  \bibinfo{author}{\bibfnamefont{A.}~\bibnamefont{Feist}},
  \bibinfo{author}{\bibfnamefont{S.}~\bibnamefont{Sch\"{a}fer}},
  \bibnamefont{and} \bibinfo{author}{\bibfnamefont{C.}~\bibnamefont{Ropers}},
  \bibinfo{journal}{Nat.\ Photon.} \textbf{\bibinfo{volume}{11}},
  \bibinfo{pages}{793} (\bibinfo{year}{2017}).

\bibitem[{\citenamefont{Pomarico et~al.}(2018)\citenamefont{Pomarico, Madan,
  Berruto, Vanacore, Wang, Kaminer, {Garc\'{\i}a de Abajo}, and
  Carbone}}]{paper306}
\bibinfo{author}{\bibfnamefont{E.}~\bibnamefont{Pomarico}},
  \bibinfo{author}{\bibfnamefont{I.}~\bibnamefont{Madan}},
  \bibinfo{author}{\bibfnamefont{G.}~\bibnamefont{Berruto}},
  \bibinfo{author}{\bibfnamefont{G.~M.} \bibnamefont{Vanacore}},
  \bibinfo{author}{\bibfnamefont{K.}~\bibnamefont{Wang}},
  \bibinfo{author}{\bibfnamefont{I.}~\bibnamefont{Kaminer}},
  \bibinfo{author}{\bibfnamefont{F.~J.} \bibnamefont{{Garc\'{\i}a de Abajo}}},
  \bibnamefont{and} \bibinfo{author}{\bibfnamefont{F.}~\bibnamefont{Carbone}},
  \bibinfo{journal}{ACS\ Photon.} \textbf{\bibinfo{volume}{5}},
  \bibinfo{pages}{759} (\bibinfo{year}{2018}).

\bibitem[{\citenamefont{Vanacore et~al.}(2018)\citenamefont{Vanacore, Madan,
  Berruto, Wang, Pomarico, Lamb, McGrouther, Kaminer, Barwick, {Garc\'{\i}a de
  Abajo} et~al.}}]{paper311}
\bibinfo{author}{\bibfnamefont{G.~M.} \bibnamefont{Vanacore}},
  \bibinfo{author}{\bibfnamefont{I.}~\bibnamefont{Madan}},
  \bibinfo{author}{\bibfnamefont{G.}~\bibnamefont{Berruto}},
  \bibinfo{author}{\bibfnamefont{K.}~\bibnamefont{Wang}},
  \bibinfo{author}{\bibfnamefont{E.}~\bibnamefont{Pomarico}},
  \bibinfo{author}{\bibfnamefont{R.~J.} \bibnamefont{Lamb}},
  \bibinfo{author}{\bibfnamefont{D.}~\bibnamefont{McGrouther}},
  \bibinfo{author}{\bibfnamefont{I.}~\bibnamefont{Kaminer}},
  \bibinfo{author}{\bibfnamefont{B.}~\bibnamefont{Barwick}},
  \bibinfo{author}{\bibfnamefont{F.~J.} \bibnamefont{{Garc\'{\i}a de Abajo}}},
  \bibnamefont{et~al.}, \bibinfo{journal}{Nat.\ Commun.}
  \textbf{\bibinfo{volume}{9}}, \bibinfo{pages}{2694} (\bibinfo{year}{2018}).

\bibitem[{\citenamefont{Cai et~al.}(2018)\citenamefont{Cai, Reinhardt, Kaminer,
  and {Garc\'{\i}a de Abajo}}}]{paper312}
\bibinfo{author}{\bibfnamefont{W.}~\bibnamefont{Cai}},
  \bibinfo{author}{\bibfnamefont{O.}~\bibnamefont{Reinhardt}},
  \bibinfo{author}{\bibfnamefont{I.}~\bibnamefont{Kaminer}}, \bibnamefont{and}
  \bibinfo{author}{\bibfnamefont{F.~J.} \bibnamefont{{Garc\'{\i}a de Abajo}}},
  \bibinfo{journal}{Phys.\ Rev.\ B} \textbf{\bibinfo{volume}{98}},
  \bibinfo{pages}{045424} (\bibinfo{year}{2018}).

\bibitem[{\citenamefont{Morimoto and Baum}(2018)}]{MB18}
\bibinfo{author}{\bibfnamefont{Y.}~\bibnamefont{Morimoto}} \bibnamefont{and}
  \bibinfo{author}{\bibfnamefont{P.}~\bibnamefont{Baum}},
  \bibinfo{journal}{Phys.\ Rev.\ A} \textbf{\bibinfo{volume}{97}},
  \bibinfo{pages}{033815} (\bibinfo{year}{2018}).

\bibitem[{\citenamefont{Tizei and Kociak}(2013)}]{TK13}
\bibinfo{author}{\bibfnamefont{L.~H.~G.} \bibnamefont{Tizei}} \bibnamefont{and}
  \bibinfo{author}{\bibfnamefont{M.}~\bibnamefont{Kociak}},
  \bibinfo{journal}{Phys.\ Rev.\ Lett.} \textbf{\bibinfo{volume}{110}},
  \bibinfo{pages}{153604} (\bibinfo{year}{2013}).

\bibitem[{\citenamefont{Meuret et~al.}(2015)\citenamefont{Meuret, Tizei,
  Cazimajou, Bourrellier, Chang, Treussart, and Kociak}}]{MTC15}
\bibinfo{author}{\bibfnamefont{S.}~\bibnamefont{Meuret}},
  \bibinfo{author}{\bibfnamefont{L.~H.~G.} \bibnamefont{Tizei}},
  \bibinfo{author}{\bibfnamefont{T.}~\bibnamefont{Cazimajou}},
  \bibinfo{author}{\bibfnamefont{R.}~\bibnamefont{Bourrellier}},
  \bibinfo{author}{\bibfnamefont{H.~C.} \bibnamefont{Chang}},
  \bibinfo{author}{\bibfnamefont{F.}~\bibnamefont{Treussart}},
  \bibnamefont{and} \bibinfo{author}{\bibfnamefont{M.}~\bibnamefont{Kociak}},
  \bibinfo{journal}{Phys.\ Rev.\ Lett.} \textbf{\bibinfo{volume}{114}},
  \bibinfo{pages}{197401} (\bibinfo{year}{2015}).

\bibitem[{\citenamefont{Bourrellier et~al.}(2016)\citenamefont{Bourrellier,
  Meuret, Tararan, St\'ephan, Kociak, Tizei, and Zobelli}}]{BMT16}
\bibinfo{author}{\bibfnamefont{R.}~\bibnamefont{Bourrellier}},
  \bibinfo{author}{\bibfnamefont{S.}~\bibnamefont{Meuret}},
  \bibinfo{author}{\bibfnamefont{A.}~\bibnamefont{Tararan}},
  \bibinfo{author}{\bibfnamefont{O.}~\bibnamefont{St\'ephan}},
  \bibinfo{author}{\bibfnamefont{M.}~\bibnamefont{Kociak}},
  \bibinfo{author}{\bibfnamefont{L.~H.~G.} \bibnamefont{Tizei}},
  \bibnamefont{and} \bibinfo{author}{\bibfnamefont{A.}~\bibnamefont{Zobelli}},
  \bibinfo{journal}{Nano\ Lett.} \textbf{\bibinfo{volume}{16}},
  \bibinfo{pages}{4317} (\bibinfo{year}{2016}).

\bibitem[{\citenamefont{Kfir}(2019)}]{K19}
\bibinfo{author}{\bibfnamefont{O.}~\bibnamefont{Kfir}},
  \bibinfo{journal}{arXiv} \textbf{\bibinfo{volume}{0}},
  \bibinfo{pages}{1902.07209v} (\bibinfo{year}{2019}).

\bibitem[{\citenamefont{{Garc\'{\i}a de Abajo}
  et~al.}(2016)\citenamefont{{Garc\'{\i}a de Abajo}, Barwick, and
  Carbone}}]{paper272}
\bibinfo{author}{\bibfnamefont{F.~J.} \bibnamefont{{Garc\'{\i}a de Abajo}}},
  \bibinfo{author}{\bibfnamefont{B.}~\bibnamefont{Barwick}}, \bibnamefont{and}
  \bibinfo{author}{\bibfnamefont{F.}~\bibnamefont{Carbone}},
  \bibinfo{journal}{Phys.\ Rev.\ B} \textbf{\bibinfo{volume}{94}},
  \bibinfo{pages}{041404(R)} (\bibinfo{year}{2016}).

\bibitem[{\citenamefont{Glauber}(1963)}]{G1963}
\bibinfo{author}{\bibfnamefont{R.~J.} \bibnamefont{Glauber}},
  \bibinfo{journal}{Phys.\ Rev.} \textbf{\bibinfo{volume}{131}},
  \bibinfo{pages}{2766} (\bibinfo{year}{1963}).

\bibitem[{\citenamefont{Lucas et~al.}(1970)\citenamefont{Lucas, Kartheuser, and
  Badro}}]{LKB1970}
\bibinfo{author}{\bibfnamefont{A.~A.} \bibnamefont{Lucas}},
  \bibinfo{author}{\bibfnamefont{E.}~\bibnamefont{Kartheuser}},
  \bibnamefont{and} \bibinfo{author}{\bibfnamefont{R.~G.} \bibnamefont{Badro}},
  \bibinfo{journal}{Phys.\ Rev.\ B} \textbf{\bibinfo{volume}{2}},
  \bibinfo{pages}{2488} (\bibinfo{year}{1970}).

\bibitem[{\citenamefont{Loudon}(2000)}]{L1983}
\bibinfo{author}{\bibfnamefont{R.}~\bibnamefont{Loudon}},
  \emph{\bibinfo{title}{The Quantum Theory of Light}}
  (\bibinfo{publisher}{Oxford University Press}, \bibinfo{address}{Oxford},
  \bibinfo{year}{2000}).

\bibitem[{\citenamefont{Das et~al.}(2019)\citenamefont{Das, Blazit, Tenc\'e,
  Zagonel, Auad, Lee, Ling, Losquin, C.~Colliex, {Garc\'{\i}a de Abajo}
  et~al.}}]{paper325}
\bibinfo{author}{\bibfnamefont{P.}~\bibnamefont{Das}},
  \bibinfo{author}{\bibfnamefont{J.~D.} \bibnamefont{Blazit}},
  \bibinfo{author}{\bibfnamefont{M.}~\bibnamefont{Tenc\'e}},
  \bibinfo{author}{\bibfnamefont{L.~F.} \bibnamefont{Zagonel}},
  \bibinfo{author}{\bibfnamefont{Y.}~\bibnamefont{Auad}},
  \bibinfo{author}{\bibfnamefont{Y.~H.} \bibnamefont{Lee}},
  \bibinfo{author}{\bibfnamefont{X.~Y.} \bibnamefont{Ling}},
  \bibinfo{author}{\bibfnamefont{A.}~\bibnamefont{Losquin}},
  \bibinfo{author}{\bibfnamefont{O.~S.} \bibnamefont{C.~Colliex}},
  \bibinfo{author}{\bibfnamefont{F.~J.} \bibnamefont{{Garc\'{\i}a de Abajo}}},
  \bibnamefont{et~al.}, \bibinfo{journal}{Ultramicroscopy}
  \textbf{\bibinfo{volume}{203}}, \bibinfo{pages}{44} (\bibinfo{year}{2019}).

\bibitem[{\citenamefont{Park and Zewail}(2012)}]{PZ12}
\bibinfo{author}{\bibfnamefont{S.~T.} \bibnamefont{Park}} \bibnamefont{and}
  \bibinfo{author}{\bibfnamefont{A.~H.} \bibnamefont{Zewail}},
  \bibinfo{journal}{J.\ Phys.\ Chem.\ A} \textbf{\bibinfo{volume}{116}},
  \bibinfo{pages}{11128} (\bibinfo{year}{2012}).

\bibitem[{\citenamefont{Eberly et~al.}(1977)\citenamefont{Eberly, Shore, {Z.
  Bia\l{}ynicka-Birula}, and {I. Bia\l{}ynicki-Birula}}}]{ESB1977}
\bibinfo{author}{\bibfnamefont{J.~H.} \bibnamefont{Eberly}},
  \bibinfo{author}{\bibfnamefont{B.}~\bibnamefont{Shore}},
  \bibinfo{author}{\bibnamefont{{Z. Bia\l{}ynicka-Birula}}}, \bibnamefont{and}
  \bibinfo{author}{\bibnamefont{{I. Bia\l{}ynicki-Birula}}},
  \bibinfo{journal}{Phys.\ Rev.\ A} \textbf{\bibinfo{volume}{16}},
  \bibinfo{pages}{2038} (\bibinfo{year}{1977}).

\bibitem[{\citenamefont{Hyun et~al.}(2008)\citenamefont{Hyun, Couillard,
  Rajendran, Liddell, and Muller}}]{HCR08}
\bibinfo{author}{\bibfnamefont{J.~K.} \bibnamefont{Hyun}},
  \bibinfo{author}{\bibfnamefont{M.}~\bibnamefont{Couillard}},
  \bibinfo{author}{\bibfnamefont{P.}~\bibnamefont{Rajendran}},
  \bibinfo{author}{\bibfnamefont{C.~M.} \bibnamefont{Liddell}},
  \bibnamefont{and} \bibinfo{author}{\bibfnamefont{D.~A.}
  \bibnamefont{Muller}}, \bibinfo{journal}{Appl.\ Phys.\ Lett.}
  \textbf{\bibinfo{volume}{93}}, \bibinfo{pages}{243106}
  (\bibinfo{year}{2008}).

\bibitem[{\citenamefont{Rejaei and Khavasi}(2015)}]{RR15}
\bibinfo{author}{\bibfnamefont{B.}~\bibnamefont{Rejaei}} \bibnamefont{and}
  \bibinfo{author}{\bibfnamefont{A.}~\bibnamefont{Khavasi}},
  \bibinfo{journal}{Rev.\ Mod.\ Phys.} \textbf{\bibinfo{volume}{87}},
  \bibinfo{pages}{1379} (\bibinfo{year}{2015}).

\bibitem[{\citenamefont{Purcell}(1946)}]{P1946}
\bibinfo{author}{\bibfnamefont{E.~M.} \bibnamefont{Purcell}},
  \bibinfo{journal}{Phys.\ Rev.} \textbf{\bibinfo{volume}{69}},
  \bibinfo{pages}{681} (\bibinfo{year}{1946}).

\bibitem[{\citenamefont{Messiah}(1966)}]{M1966}
\bibinfo{author}{\bibfnamefont{A.}~\bibnamefont{Messiah}},
  \emph{\bibinfo{title}{Quantum Mechanics}}
  (\bibinfo{publisher}{North-Holland}, \bibinfo{address}{New York},
  \bibinfo{year}{1966}).

\bibitem[{\citenamefont{Sakurai}(1994)}]{S1994}
\bibinfo{author}{\bibfnamefont{J.~J.} \bibnamefont{Sakurai}},
  \emph{\bibinfo{title}{Modern Quantum Mechanics}}
  (\bibinfo{publisher}{Addison-Wesley}, \bibinfo{year}{1994}).

\bibitem[{\citenamefont{Milonni}(1994)}]{M94}
\bibinfo{author}{\bibfnamefont{P.~W.} \bibnamefont{Milonni}},
  \emph{\bibinfo{title}{The quantum vacuum: an introduction to quantum
  electrodynamics}} (\bibinfo{publisher}{Academic Press}, \bibinfo{address}{San
  Diego}, \bibinfo{year}{1994}).

\bibitem[{\citenamefont{Carruthers and Nieto}(1965)}]{CN1965}
\bibinfo{author}{\bibfnamefont{P.}~\bibnamefont{Carruthers}} \bibnamefont{and}
  \bibinfo{author}{\bibfnamefont{M.~M.} \bibnamefont{Nieto}},
  \bibinfo{journal}{Am.\ J.\ Phys.} \textbf{\bibinfo{volume}{33}},
  \bibinfo{pages}{537} (\bibinfo{year}{1965}).

\bibitem[{\citenamefont{Abrikosov et~al.}(1965)\citenamefont{Abrikosov,
  {Gor'kov}, and Dzyaloshinskii}}]{AGD1965}
\bibinfo{author}{\bibfnamefont{A.~A.} \bibnamefont{Abrikosov}},
  \bibinfo{author}{\bibfnamefont{L.~P.} \bibnamefont{{Gor'kov}}},
  \bibnamefont{and} \bibinfo{author}{\bibfnamefont{I.~Y.}
  \bibnamefont{Dzyaloshinskii}}, \emph{\bibinfo{title}{Quantum Field
  Theoretical Methods in Statistical Physics}} (\bibinfo{publisher}{Pergamon
  Press}, \bibinfo{address}{New York}, \bibinfo{year}{1965}).

\bibitem[{\citenamefont{Abramowitz and Stegun}(1972)}]{AS1972}
\bibinfo{author}{\bibfnamefont{M.}~\bibnamefont{Abramowitz}} \bibnamefont{and}
  \bibinfo{author}{\bibfnamefont{I.~A.} \bibnamefont{Stegun}},
  \emph{\bibinfo{title}{Handbook of Mathematical Functions}}
  (\bibinfo{publisher}{Dover}, \bibinfo{address}{New York},
  \bibinfo{year}{1972}).

\bibitem[{\citenamefont{Feller}(1968)}]{F1968}
\bibinfo{author}{\bibfnamefont{W.}~\bibnamefont{Feller}},
  \emph{\bibinfo{title}{An Introduction to Probability Theory and Its
  Applications}} (\bibinfo{publisher}{John Wiley}, \bibinfo{address}{New York},
  \bibinfo{year}{1968}).

\bibitem[{\citenamefont{Gradshteyn and Ryzhik}(2007)}]{GR1980}
\bibinfo{author}{\bibfnamefont{I.~S.} \bibnamefont{Gradshteyn}}
  \bibnamefont{and} \bibinfo{author}{\bibfnamefont{I.~M.}
  \bibnamefont{Ryzhik}}, \emph{\bibinfo{title}{Table of Integrals, Series, and
  Products}} (\bibinfo{publisher}{Academic Press}, \bibinfo{address}{London},
  \bibinfo{year}{2007}).

\bibitem[{\citenamefont{Scully and Zubairy}(1997)}]{SZ97}
\bibinfo{author}{\bibfnamefont{M.~O.} \bibnamefont{Scully}} \bibnamefont{and}
  \bibinfo{author}{\bibfnamefont{M.~S.} \bibnamefont{Zubairy}},
  \emph{\bibinfo{title}{Quantum Optics}} (\bibinfo{publisher}{Cambridge
  University Press}, \bibinfo{address}{Cambridge}, \bibinfo{year}{1997}).

\bibitem[{\citenamefont{{Garc\'{\i}a de Abajo}}(2013)}]{paper228}
\bibinfo{author}{\bibfnamefont{F.~J.} \bibnamefont{{Garc\'{\i}a de Abajo}}},
  \bibinfo{journal}{ACS\ Nano} \textbf{\bibinfo{volume}{7}},
  \bibinfo{pages}{11409} (\bibinfo{year}{2013}).

\bibitem[{\citenamefont{Pines and Nozi\`{e}res}(1966)}]{PN1966}
\bibinfo{author}{\bibfnamefont{D.}~\bibnamefont{Pines}} \bibnamefont{and}
  \bibinfo{author}{\bibfnamefont{P.}~\bibnamefont{Nozi\`{e}res}},
  \emph{\bibinfo{title}{The Theory of Quantum Liquids}} (\bibinfo{publisher}{W.
  A. Benjamin, Inc.}, \bibinfo{address}{New York}, \bibinfo{year}{1966}).

\bibitem[{\citenamefont{Johnson and Christy}(1972)}]{JC1972}
\bibinfo{author}{\bibfnamefont{P.~B.} \bibnamefont{Johnson}} \bibnamefont{and}
  \bibinfo{author}{\bibfnamefont{R.~W.} \bibnamefont{Christy}},
  \bibinfo{journal}{Phys.\ Rev.\ B} \textbf{\bibinfo{volume}{6}},
  \bibinfo{pages}{4370} (\bibinfo{year}{1972}).

\bibitem[{\citenamefont{Dias and {Garc\'{\i}a de Abajo}}(2019)}]{paper331}
\bibinfo{author}{\bibfnamefont{E.~J.~C.} \bibnamefont{Dias}} \bibnamefont{and}
  \bibinfo{author}{\bibfnamefont{F.~J.} \bibnamefont{{Garc\'{\i}a de Abajo}}},
  \bibinfo{journal}{ACS\ Nano} \textbf{\bibinfo{volume}{13}},
  \bibinfo{pages}{5184} (\bibinfo{year}{2019}).

\bibitem[{\citenamefont{Glauber and Lewenstein}(1991)}]{GL91}
\bibinfo{author}{\bibfnamefont{R.~J.} \bibnamefont{Glauber}} \bibnamefont{and}
  \bibinfo{author}{\bibfnamefont{M.}~\bibnamefont{Lewenstein}},
  \bibinfo{journal}{Phys.\ Rev.\ A} \textbf{\bibinfo{volume}{43}},
  \bibinfo{pages}{467} (\bibinfo{year}{1991}).

\bibitem[{\citenamefont{{Garc\'{\i}a de Abajo} and Kociak}(2008)}]{paper102}
\bibinfo{author}{\bibfnamefont{F.~J.} \bibnamefont{{Garc\'{\i}a de Abajo}}}
  \bibnamefont{and} \bibinfo{author}{\bibfnamefont{M.}~\bibnamefont{Kociak}},
  \bibinfo{journal}{Phys.\ Rev.\ Lett.} \textbf{\bibinfo{volume}{100}},
  \bibinfo{pages}{106804} (\bibinfo{year}{2008}).

\end{thebibliography}

\end{document}